\documentclass[review,3p,12pt,times]{elsarticle}

\usepackage{bm}
\newcommand{\ea}{\bm{e}_\alpha}

\newcommand{\faeq}{f_\alpha^{\rm eq}}
\newcommand{\fa}{f_\alpha}

\newcommand{\haieq}{h_\alpha^{i, \rm eq}}
\newcommand{\hai}{h_\alpha^{i}}
\newcommand{\haibareq}{\bar{h}_\alpha^{i,\rm eq}}
\newcommand{\haibar}{\bar{h}_\alpha^{i}}

\newcommand{\gaeq}{g_\alpha^{\rm eq}}
\newcommand{\ga}{g_\alpha}
\newcommand{\gabareq}{\bar{g}_\alpha^{\rm eq}}
\newcommand{\gabar}{\bar{g}_\alpha}
\newcommand{\Gaa}{\varGamma_\alpha}
\newcommand{\etal}{\textit{et al.} }

\newcommand{\ie}  {\textit{i.e.}}

\usepackage{multicol}
\usepackage{etoolbox}  
\usepackage{framed} 
\usepackage{gensymb}
\usepackage{subcaption}
\usepackage{epstopdf}

\usepackage{setspace}
\usepackage{float}

\usepackage{sectsty}
\sectionfont{\fontsize{15}{15}\selectfont}

\usepackage{lipsum}  
\usepackage[T1]{fontenc}
\usepackage{polski}
\usepackage[utf8]{inputenc}
\usepackage[english]{babel}
\usepackage{amsmath, amsthm, amsfonts, amssymb}
\usepackage{bbm}

\usepackage{hyperref} 
\hypersetup{urlcolor=blue, citecolor=blue, filecolor=blue, linkcolor=blue,urlcolor=blue,colorlinks=true}

\usepackage{pgfplots}
\usepackage{pgfplotstable}
\usepackage{tikz} 
\usetikzlibrary{positioning,patterns}

\usepackage{multirow}

\usepackage[abs]{overpic}
\usepackage{lineno}
\usepackage{comment}
\usepackage{graphicx}
\usepackage[tableposition=top]{caption}
\usepackage{longtable}

\usepackage[capitalise]{cleveref}

\definecolor{niceblue}{rgb}{0.121,0.231,0.4}
\definecolor{nicered}{rgb}{0.76,0.3,0.19}

\begin{document}

\large
\begin{frontmatter}
\title{Drag coefficient of a rising bubble in a shear-thinning fluid using the power-law scheme coupled with a Cahn-Hilliard equation with a variable mobility: A lattice Boltzmann study and comparison with experiment}

\author{Amirabbas Ghorbanpour Arani}
\author{Reza Haghani-Hassan-Abadi}
\author{Mohammad Majidi}
\author{Mohammad-Hassan Rahimian\corref{cor1}}
\ead{rahimyan@ut.ac.ir}
\cortext[cor1]{Corresponding author}

\address{School of Mechanical Engineering, College of Engineering, University of Tehran, Tehran, Iran}

\begin{abstract}
This study aims to investigate the behavior of multicomponent fluid flows consisting of Newtonian and non-Newtonian components, especially terminal velocity of a rising bubble in a power-law fluid. A recent lattice Boltzmann (LB) model is extended using power-law scheme to be able to simulate both Newtonian and non-Newtonian fluid flows at high density and viscosity ratios. Also, a variable mobility is introduced in this study to minimize the unphysical error around small bubbles in the domain. A three-component fluid flow system is examined using a constant and variable mobility. It is shown that each component has more stability using variable mobility while constant mobility causes interface dissipation, leading to mass loss gradually. In addition, two test cases including power-law fluid flows driven between two parallel plates are conducted to show the accuracy and capability of the model. To find a grid-independent computational domain, a grid independency test is carried out to show that a $200\times 400$ domain size is suitable for our computations. Then, terminal velocity of a rising bubble is compared to an existing correlation in the literature, indicating that the results are in good agreement with existing study so that average relative error in six different cases is 5.66 $\%$. Also, the simulated examples show good conformity to experimental results over a range of the Reynolds and E\"{o}tv\"{o}s numbers.
\end{abstract}


\end{frontmatter}


\section{Introduction}

Multiphase flows, especially rising bubble in Newtonian and non-Newtonian liquids, are widely applicable in the industry. Multiphase flow systems exist in many contexts such as water treatment processes, biomedical engineering, pharmaceutical and food industries, and so on \cite{chhabra2006bubbles, chhabra2011non}. Also, multiphase flow systems are found in the droplet-based microfluidic devices which is applicable in drug delivery and cell analysis \cite{PhysRevE.102.023114, farhadi2021, wang}.  Therefore, many research efforts are targeted on predicting interaction between liquid and gas not only in Newtonian liquids but also in non-Newtonian ones. Since gas phase is in the form of bubble in many systems, it is of importance to find the behaviors and characteristics of the bubble in Newtonian and non-Newtonian liquids. As a matter of fact, understanding two-phase flow behavior can help improve and maintain such systems. Terminal velocity of a bubble is an important parameter in two-phase flow systems which plays an important role in optimization and designing of gas-liquid systems \cite{BATTISTELLA2020104249}, showing the necessity of investigating bubble dynamics in Newtonian and non-Newtonian liquids. In addition, terminal velocity can exhibit equilibrium in a system, as it occurs when drag and buoyant force are balanced. 

Numerical analysis has become very popular because of its prediction and comprehension of phenomena in nature. Many numerical approaches have been applied to analyze multiphase flows in macroscopic scale like front-tracking \cite{TRYGGVASON2001708}, volume of fluid \cite{HIRT1981201}, level set method \cite{PILLAPAKKAM2001552}, etc. Battistella \etal \cite{BATTISTELLA2020104249} used a front-tracking model to simulate rising bubble in a power-law fluid. They used a drag closure proposed by dijkhuizen \etal \cite{20cad03008e541e78c2c7e51aa97e6e7} to obtain terminal velocity of bubble, and compare it with numerical simulation results. In most cases, the accuracy of the model was 20$\%$. Li \etal \cite{lima} experimentally investigated rising bubble in a non-Newtonian liquid. They proposed a drag coefficient correlation by considering the Reynolds number, aspect ratio, and rheological properties of the liquid. Also, they plotted a map of shape regimes for bubbles in a non-Newtonian fluid using Reynolds, E\"{o}tv\"{o}s, and Morton numbers. Islam \etal \cite{islammd} used volume-of-fluid (VOF) approach to investigate rising bubble behavior in a Newtonian and non-Newtonian liquid. Behavior of a non-Newtonian liquid was described by power-law equation. They studied terminal velocity and deformation shape of bubble using numerical simulations. They showed that their results are in good agreement with empirical correlations. Ohta \etal \cite{Ohta_2009} used coupled level-set and volume-of-fluid (CLSVOF) to simulate bubble rising in a non-Newtonian liquid. The viscosity of liquid was determined by Carreau model. They investigated bubble rise velocity and bubble shape in non-Newtonian fluids by changing Carreau model parameters. They concluded that in case of implementing effective parameters, Bhaga-Weber map can predict bubble shape in the non-Newtonian fluid. Premlata \etal \cite{PREMLATA201753} carried out VOF method using  the Carreau-Yasuda model to simulate rising bubble in a non-Newtonian fluid. They validated their model by comparing with results in the literature. They showed that rising velocity increases by increasing shear-thinning tendency and deformation of bubble increases at high Gallilei number and low E\"{o}tv\"{o}s number. Pang and Lu \cite{PANG2018101} used VOF method to simulate bubble rising in Carreau fluid. They investigated the influence of surface tension, rheological index, and different gravity field on bubble dynamics. They found that bubble deformation increases by increasing Gallilei and E\"{o}tv\"{o}s numbers and decreases with the increase of rheological index. Also, they showed that terminal velocity increases by increasing Gallilei number and decreasing rheological index.

In the recent decades, lattice Boltzmann method (LBM) has been used as an alternative way to simulate fluids dynamics in mesoscopic scale. The LBM has several advantages over traditional computational fluid dynamics (CFD) methods such as ease of implementation, capability of scaling on parallel computers, and handling complex geometries \cite{kruger2017lattice, MAJIDI2022103846, farhadi2022marangoni}. Several LB models have been proposed for two-phase fluids such as color gradient \cite{PhysRevA.43.4320}, pseudopotential \cite{PhysRevE.47.1815}, free energy \cite{PhysRevE.54.5041}, and mean-field \cite{HE1999642}. All of these early models confront with some drawbacks like limited density ratio and spurious currents (non-physical velocity fields). Inamuro \etal \cite{INAMURO2004628} proposed a model based on free energy that can capture large density ratio, but the Poisson equation in the model reduced simplicity and efficiency of the LBM. Lee \cite{LEE2009987} proposed a two-phase model using Cahn-Hilliard (CH) interface equation that can handle large density and viscosity ratios. Also, spurious currents is in order of $10^{-14}$ which are negligible. Haghani-Hassan-Abadi \etal \cite{hagh} developed Lee's two-phase model to three-component model, implementing bulk free energy and CH equation proposed in \cite{refId0}. Their model maintained characteristics of Lee's binary model and can be used for one, two, and three component fluids. This model was validated by several cases and capable of analyzing phenomena in nature and industry \cite{ghorbanpp}. Beside above models, many efforts have been done to capture non-Newtonian behavior of fluids in LBM. Gabbanelli \etal \cite{gabbna} extended LBM to non-Newtonian formulation using a truncated power-law model to capture non-Newtonian characteristics of single-component fluids. Boyd \etal \cite{Boyd_2006} proposed a second order LB model to simulate non-Newtonian fluids. Their model is efficient computationally and more accurate compared to previous models. Moreover, further works were done for multiphase non-Newtonian fluids \cite{regui7, haihuliu, XIE2016118}.

To the authors' best knowledge, the model introduced in \cite{hagh} is strong due to many numerical tests conducted by the authors. Therefore, in this study, this model is used for the computation of multiphase flows. In this paper, we enhance the model using variable mobility to minimize little numerical dissipation around small bubbles reported in \cite{ghorbanpp}. Also, Variable and constant mobilities are compared to show how mass conservation is satisfied in the present study. Variable mobility can effectively reduce dissolutions of bubble and droplets to surrounding fluid \cite{refId0, voluchi,CiCP-7-362}. Also, we couple the mentioned model to the power-law scheme, using the truncated power-law model for viscosity distribution, to simulate non-Newtonian fluids at high density and viscosity ratios.

The paper is organized as follows. Section 2 represents lattice Boltzmann equations (LBEs) together with CH equations and explanations about the truncated power-law model. Benefits of variable mobility and validation of method, which includes two non-Newtonian test cases studied by the present method and compared with analytical solution to show accuracy and robustness of the extension, are carried out in section 3. Then in section 4, grid convergence test is carried out and rising behavior of a bubble in a shear-thinning fluid is examined. Also, terminal velocity and deformation shape of the bubble are compared with empirical results. Finally, this work is concluded in section 5.

\section{Methodology}
\subsection{Interface capturing model}

Consider a domain $\varOmega$ of three incompressible, immiscible fluids. Each fluid is distinguished by its phase-field variable ($C_i \quad i=1-3$) along with the following constraint:

\begin{eqnarray} \label{eq:constraint}
C_1+C_2+C_3=1
\end{eqnarray}
Free energy for three-component flows is as follows \cite{refId0}:
\begin{eqnarray} \label{eq:freeenergy}
F(C_1,C_2,C_3)=\int_{\varOmega}^{} \left[ \frac{12}{\xi} E_0(C_1,C_2,C_3 )+\frac{3}{8} \xi \Sigma_1 \left| \bm{\nabla} C_1 \right|^2+\frac{3}{8} \xi \Sigma_2 \left| \bm{\nabla} C_2 \right|^2+\frac{3}{8} \xi \Sigma_3 \left| \bm{\nabla} C_3 \right|^2\right] d\varOmega
\end{eqnarray}
where $\xi$ is the interface thickness which is usually of 3 to 5 lattices, $\Sigma_i$ are capillary coefficients, and $E_0$ is the bulk free energy. For consistency with diphasic systems, the capillary coefficients are defined as $\Sigma_i=\sigma_{ij}+\sigma_{ik}-\sigma_{jk} \quad (i,j,k=1-3)$ where $\sigma_{ij}$ is the surface tension coefficient between fluid $i$ and fluid $j$. The bulk free energy is defined as
\begin{eqnarray} \label{eq:bulkfreeenergy}
E_0=\sigma_{12} C_1^2 C_2^2+\sigma_{13} C_1^2 C_3^2+\sigma_{23} C_2^2 C_3^2+C_1 C_2 C_3 (\Sigma_1 C_1+\Sigma_2 C_2+\Sigma_3 C_3)+\varLambda C_1^2 C_2^2 C_3^2
\end{eqnarray}
where $\varLambda$ is a positive parameter for total spreading scenarios. 

The three-component CH equations can be written as following: 
\begin{eqnarray} \label{eq:chequation}
\dfrac{\partial C_i}{\partial t}+\bm{\nabla} \cdot (\bm{u} C_i) =\bm{\nabla} \cdot (D_i \bm{\nabla} \mu_i)  \quad i=1-3
\end{eqnarray}

\begin{eqnarray}
\mu_i=\frac{4 \Sigma_T}{{\xi}} \sum_{j\neq i} \dfrac{1}{\Sigma_{j}}[\partial_i E_0 -\partial_j E_0 ]-\dfrac{3}{4}\xi \Sigma_{i} \nabla^2 C_i  \quad i=1-3
\end{eqnarray}
where $t$ is the time, $\bm{u}$ is the velocity, and $\mu_i$ is the chemical potential of the fluid $i$.  the mobility of the fluid $i$, $D_i$, and  $\Sigma_{T}$ are defined as
\begin{equation} \label{eq:mobility} 
D_i = \left\{ \begin{array}{rcl}
\frac{D_0}{\Sigma_i} 
& Constant \ Mobility& \\
\frac{D_0}{\Sigma_i}C_i(1-C_i) 
& Variable \ Mobility
\end{array}\right. \quad i=1-3
\end{equation}
\begin{eqnarray}
\frac{3}{\Sigma_{T}}=\frac{1}{\Sigma_{1}}+\frac{1}{\Sigma_{2}}+\frac{1}{\Sigma_{3}}
\end{eqnarray}	
where $D_{0}=\xi/300$ is an auxiliary parameter \cite{hagh}.

\subsection{Lattice Boltzmann equations}
The LBE with the single relaxation time can be written as \cite{LEE2009987} 
\begin{equation} \label{eq:LBEf}
\fa(\bm{x}+\ea\delta t,t+\delta t)-\fa(\bm{x},t)=-\dfrac{1}{\lambda}(\fa-\faeq)+\dfrac{ (\ea-\bm{u})\cdot \bm{F}\faeq}{\rho c_s^2}
\end{equation}
where $\fa$ is the particle distribution function in the $\alpha$-direction, $\lambda$ is the relaxation time, $\rho$ is the density of mixture, and $\bm{F}$ is the intermolecular force modeling non-ideal gas effects.  $\faeq$ is the equilibrium distribution function which are defined as follows:
\begin{equation}
\faeq=w_\alpha \rho \left[ 1+\frac{(\ea \cdot \bm{u})}{c_s^2}+\frac{(\ea \cdot \bm{u})^2}{2c_s^4}-\frac{(\bm{u} \cdot \bm{u})}{2c_s^2} \right]
\label{Gamma}
\end{equation}

In this study, the D2Q9 lattice structure is used with the following weight coefficients, $w_\alpha$, and microscopic velocity, $\ea$, set:
\begin{equation}
w_\alpha =
\begin{cases} 
4/ 9, & \alpha=0 \\
1/ 9, & \alpha=1-4 \\
1/36, & \alpha=5-8
\end{cases}
\label{weight}
\end{equation}
\begin{equation}
\ea=c
\begin{cases} 
(0,0), & \alpha=0 \\
(\cos[(\alpha-1)\pi /2],\sin[(\alpha-1)\pi/2]), & \alpha=1-4 \\
(\cos[(2\alpha-9)\pi /4],\sin[(2\alpha-9)\pi/4])\sqrt{2}, & \alpha=5-8
\end{cases}
\label{alpha}
\end{equation}
In the above equations, $c_s=\dfrac{c}{\sqrt{3}}$ is the speed of sound in the D2Q9 lattice in which $c=\delta x/\delta t$ is the lattice speed with $\delta x$ and $\delta t$ being the length scale and time scale, respectively. 
The force term $\bm{F}$ is expressed as
\begin{equation} \label{eq:force}
	\bm{F}=\bm{\nabla}\rho c_s^2-\bm{\nabla}p+\sum_{i=1}^{3}\mu_i \bm{\nabla} C_i +\bm{F}_b
\end{equation}
where $\bm{F}_b=-\rho g \overrightarrow{j}$ is the body force with $g$ being the gravitational acceleration. 

To obtain a LBE for recovering the hydrodynamics properties of the mixture, the following transformation is used:

\begin{equation} \label{eq:gdistributionf}
\ga=\fa c_s^2 +(p-\rho c_s^2)\varGamma_\alpha (0)
\end{equation}
where $\varGamma_\alpha(\bm{u})=\faeq/\rho$. Taking the total derivative of $\ga$ and using the trapezoidal rule results in the following LBE:
\begin{equation} \label{eq:LBEg}
\begin{aligned}
\gabar(\bm{x}+\ea\delta t,t+\delta t)-\gabar(\bm{x},t) \\
&=-\frac{1}{\tau +0.5}(\gabar-\gabareq)\rvert_{(\bm{x},t)} \\
&+\delta t(\ea-\bm{u})\cdot\left[\bm{\nabla} \rho c_s^2(\Gaa-\Gaa(0))+(\sum_{i=1}^{3}\mu_i \bm{\nabla} C_i +\bm{F}_b)\Gaa\right]_{(\bm{x},t)}
\end{aligned}
\end{equation}
where the dimensionless relaxation time $\tau=\lambda/\delta t$ is related to kinematic viscosity $\nu$ by $\tau=\nu/c_s^2$. In the above equations, where $\gabar$ and $\gabareq$ are modified particle and equilibrium distribution functions, respectively, which are defined as:
\begin{equation} \label{eq:gbar}
\begin{split}
&\gabar=\ga+\frac{1}{2\tau}(\ga-\gaeq)-\frac{\delta t}{2}(\ea-\bm{u})\cdot\left[\bm{\nabla} \rho c_s^2(\Gaa-\Gaa(0))+(\sum_{i=1}^{3}\mu_i \bm{\nabla} C_i +\bm{F}_b)\Gaa\right] 
\end{split}
\end{equation}
\begin{equation} \label{eq:gbareq}
\begin{aligned}
& \gabareq=\gaeq-\frac{\delta t}{2}(\ea-\bm{u})\cdot\left[\bm{\nabla} \rho c_s^2(\Gaa-\Gaa(0))+(\sum_{i=1}^{3}\mu_i \bm{\nabla} C_i +\bm{F}_b)\Gaa\right]
\end{aligned}
\end{equation}

In order to track the interfaces among three fluids, we use the following distribution functions 
\begin{equation} \label{LBEh}
\hai=\frac{C_i}{\rho}\fa    \quad i=1,2
\end{equation}

Note that $C_3$ is obtained via the constraint in Eq.~\eqref{eq:constraint}.

Like previously, by taking the total derivative of $\hai$ and using the trapezoidal rule, we have the following LBE for interfaces capturing:
\begin{equation} \label{eq:interface}
\begin{split}
& \haibar(\bm{x}+\ea \delta t, t+\delta t)-\haibar(\bm{x},t)=-\frac{1}{\tau_h+0.5}(\haibar-\haibareq)\rvert_{(\bm{x},t)} \\
&+\delta t(\ea -\bm{u})\cdot \left[\bm{\nabla} C_i-\frac{C_i}{\rho  c_s^2}(\bm{\nabla} p-\sum_{i=1}^{3}\mu_i \bm{\nabla} C_i-\bm{F}_b)\right]\Gaa \rvert_{(\bm{x},t)}\\
&  +\frac{\delta t}{2} D_i \nabla^2 \mu_i\Gaa\rvert_{(\bm{x},t)} 
+ \frac{\delta t}{2} D_i \nabla^2 \mu_i\Gaa\rvert_{(\bm{x}+\ea \delta t,t)} \\
&+\frac{\delta t}{2} \frac{D_i}{C_i(1-C_i)} \bm{\nabla}\mu_i\Gaa[\bm{\nabla} C_i-\bm{\nabla}C_i^2 ]\rvert_{(\bm{x},t)}+\frac{\delta t}{2} \frac{D_i}{C_i(1-C_i)} \bm{\nabla}{\mu}_i\Gaa[\bm{\nabla} C_i-\bm{\nabla}C_i^2 ]\rvert_{(\bm{x}+\ea\delta t,t)} \\
& \quad i=1,2
\end{split}
\end{equation}
where $\tau_h=0.5$ is the dimensionless relaxation time. $\haibar$ and $\haibareq$ are the modified particle and equilibrium distribution functions, respectively, defined as 
\begin{equation}
\begin{aligned}
&\haibar=\hai+\frac{1}{2\tau}(\hai-\haieq)-\frac{\delta t}{2}(\ea -\bm{u})\cdot \left[\bm{\nabla} C_i-\frac{C_i}{\rho  c_s^2}(\bm{\nabla} p-\sum_{i=1}^{3}\mu_i \bm{\nabla} C_i-\bm{F}_b)\right]\Gaa
\end{aligned}
\end{equation}

\begin{equation}
\begin{split}
&\haibareq=\haieq-\frac{\delta t}{2}(\ea -\bm{u})\cdot \left[\bm{\nabla} C_i-\frac{C_i}{\rho  c_s^2}(\bm{\nabla} p-\sum_{i=1}^{3}\mu_i \bm{\nabla} C_i-\bm{F}_b)\right]\Gaa
\end{split}
\end{equation}
where $\haieq=C_i \Gaa$.

As mentioned earlier, in this study we employ a variable mobility as it shows better mass conservation properties. Equation~\eqref{eq:mobility} defines both constant and variable mobilities. As can be seen, the variable mobility is not constant and dependent of $C_i$. As such, when taking the total derivative of Eq.~\eqref{LBEh} and considering the variable mobilities, the diffusion term on the right-hand side of Eq.~\eqref{eq:chequation} simplifies to
\begin{equation}
\begin{split}
&\bm{\nabla}\cdot(D_i\bm{\nabla} \mu_i) =\bm{\nabla}\cdot [\frac{D_0}{\Sigma_i} C_i(1-C_i)  \bm{\nabla}\mu_i]\\
&=\frac{D_0}{\Sigma_i} C_i(1-C_i) \nabla^2 \mu_i+\frac{D_0}{\Sigma_i} (\bm{\nabla} C_i-\bm{\nabla}C_i^2)\bm{\nabla} \mu_i \\
& =D_i \nabla^2 \mu_i + \frac{D_i}{C_i(1-C_i)} (\bm{\nabla} C_i-\bm{\nabla}C_i^2)\bm{\nabla} \mu_i \quad i=1,2
\end{split}
\end{equation}
which the last term is related to the variable mobilities and also appears in the interface capturing LBE, \ie, Eq.~\eqref{eq:interface}. 

In the above equations, second-order central and mixed finite differences and directional derivatives of a scaler variable are implemented to enhance the numerical stability \cite{ghorbanpp}.

The phase-field variables, momentum and pressure can be calculated by taking the moments of the modified distribution functions 
\begin{equation}
C_i=\sum_{\alpha}^{}\haibar \quad i=1,2
\end{equation}
\begin{equation}
\rho\bm{u}=\frac{1}{c_s^2}\sum_{\alpha}^{}\ea \gabar+\frac{\delta t}{2}(\sum_{i=1}^{3}\mu_i \bm{\nabla} C_i+\bm{F}_b)
\end{equation}
\begin{equation}
p=\sum_{\alpha}^{}\gabar+\bm{u} \cdot \bm{\nabla} \rho c_s^2
\end{equation}


\subsection{Strain rate tensor and relaxation time calculation}

In the continuum fluid flows, the Navier-stokes (NS) equations are recovered from LBE through the Chapman-Enskog (CE) analysis. The strain rate tensor is calculated by the second moment of non-equilibrium distribution function as follows \cite{philips, hosseini2022lattice}:
\begin{eqnarray} \label{eq:hydrodynamics}
    D = \frac{1}{2}(\bm{\nabla} \bm{u} + \nabla {\bm{u}}^t)=-\frac{\Pi^{(1)}}{2(\tau+0.5)\rho {c_s}^2} 
\end{eqnarray}
where $\bm{u}^t$ denotes the transposed matrix to $\bm{u}$ and $\Pi^{(1)}$ as first-order momentum flux for the D2Q9 lattice is given as

\begin{equation} \label{eq:momentum}
 \Pi^{(1)}_{\alpha \beta}(\bm{x},t) = \sum_{i= 0}^{8} f^{(1)}_i (\bm{x},t)e_{\alpha i}e_{\beta i}
\end{equation}
where $f_i^{(1)}$ is a first-order perturbation distribution function \cite{philips}. Using the transformation in Eq.~\eqref{eq:gdistributionf}, the relationship between non-equilibrium particle distribution functions can be obtained as follows:
\begin{equation} 
\ga - \gaeq = (\fa-\faeq)c^2_s
\end{equation}
By using Eqs.~\eqref{eq:gbar} and~\eqref{eq:gbareq} we have
\begin{equation}
    \ga - \gaeq = \frac{2\tau}{2\tau + 1}(\gabar - \gabareq)
\end{equation}
As such, the non-equilibrium particle distribution function of $f$ can be replaced as follows:
\begin{equation}
    \fa-\faeq = \frac{1}{c^2_s}\frac{2\tau}{2\tau + 1}(\gabar -\gabareq)
\end{equation}
As a result, knowing the magnitude of non-equilibrium distribution function of $f$, $\fa-\faeq$, Eq. \eqref{eq:momentum} and Eq. \eqref{eq:hydrodynamics} can be solved at each time step to obtain strain rate tensor, \ie, $D$.

Constitutive equation for a generalized Newtonian fluid is defined as \cite{irgens2014rheology}
\begin{equation} \label{eq:T}
    T=2\eta(\dot{\gamma})D
\end{equation}
where $\eta(\dot{\gamma}) $ and $T$ are the apparent viscosity function of shear rate and shear stress tensor, respectively. The non-Newtonian fluid in this study is modeled as a power-law fluid with consistency $k$ and index $n$. Therefore, the apparent viscosity in Eq.~\eqref{eq:T} is defined as follows:
\begin{equation} \label{eq:apparentviscosity}
    \eta(\dot{\gamma}) = k\dot{\gamma}^{n-1}
\end{equation}
 The shear rate, as an invariant of strain rate tensor, is expressed as $\dot\gamma=\sqrt{2D:D}$, and can be calculated locally by the non-equilibrium particle distribution function using Eq. \eqref{eq:hydrodynamics}.
 
 In the non-Newtonian fluids, viscosity is not constant, so relaxation parameter is calculated at each lattice site by recalling the relationship between fluid viscosity and relaxation parameter:
\begin{equation} \label{eq:ref}
    \tau=\frac{ \eta({\dot{\gamma}}(\tau))}{\rho c^2_s}
\end{equation}
Therefore, the relaxation parameter obtained by Eq. \eqref{eq:ref} is used in Eq.\eqref{eq:LBEg}.The above equation is an implicit equation which can be solved by iterative method numerically. However, in this study, $\dot\gamma(\tau)$  is updated by the relaxation parameter at the previous time step \cite{philips,malasp}. Such approach uses much less computational cost.
 
Apparent viscosity corresponds to shear-thinning fluid, shear-thickening fluid, and Newtonian fluid when $n$ is less than 1, more than 1, and equal to 1, respectively. Therefore, when shear rate increases, the viscosity of shear-thinning fluid decreases. As such, the shear rate in this study is computed by particle distribution function rather than using derivatives of velocities by finite difference scheme. This approach, deriving parameters locally, is more consistent with the philosophy of the LBM \cite{philips}.

Direct implementation of a power-law fluid flow in the LBM confronts an important problem. In a shear-thinning fluid ($n<1$) and at zero shear rates ($\dot\gamma=0$), the relaxation parameter diverges. On the other hand, in a shear-thickening fluid ($n>1$) and at zero shear rates ($\dot\gamma=0$), the relaxation parameter becomes zero, leading to instability in solution as well. Hence, in this study the truncated power-law viscosity model is implemented to limit the range of viscosity as follows \cite{gabbna}:
\begin{equation}
 \eta(\dot{\gamma}) = \left\{ \begin{array}{rcl}
k\dot{\gamma}^{n-1}_0 & \mbox{for}
& \dot{\gamma}<\dot{\gamma}_0 \\ k\dot{\gamma}^{n-1} & \mbox{for} &  \dot{\gamma}_0<\dot{\gamma}<\dot{\gamma}_{\infty} \\
k\dot{\gamma}^{n-1}_{\infty} & \mbox{for} & \dot{\gamma}_{\infty}<\dot{\gamma}
\end{array}\right.
\end{equation}
Therefore, in the truncated power-law model, range of relaxation parameter is limited and prevents solution from instability.

\section{Validation}
\subsection{Effect of variable mobility}
\begin{figure}[H]
    \centering
   \includegraphics[width= 1\textwidth,trim={0cm 0cm 0cm 0cm},clip]{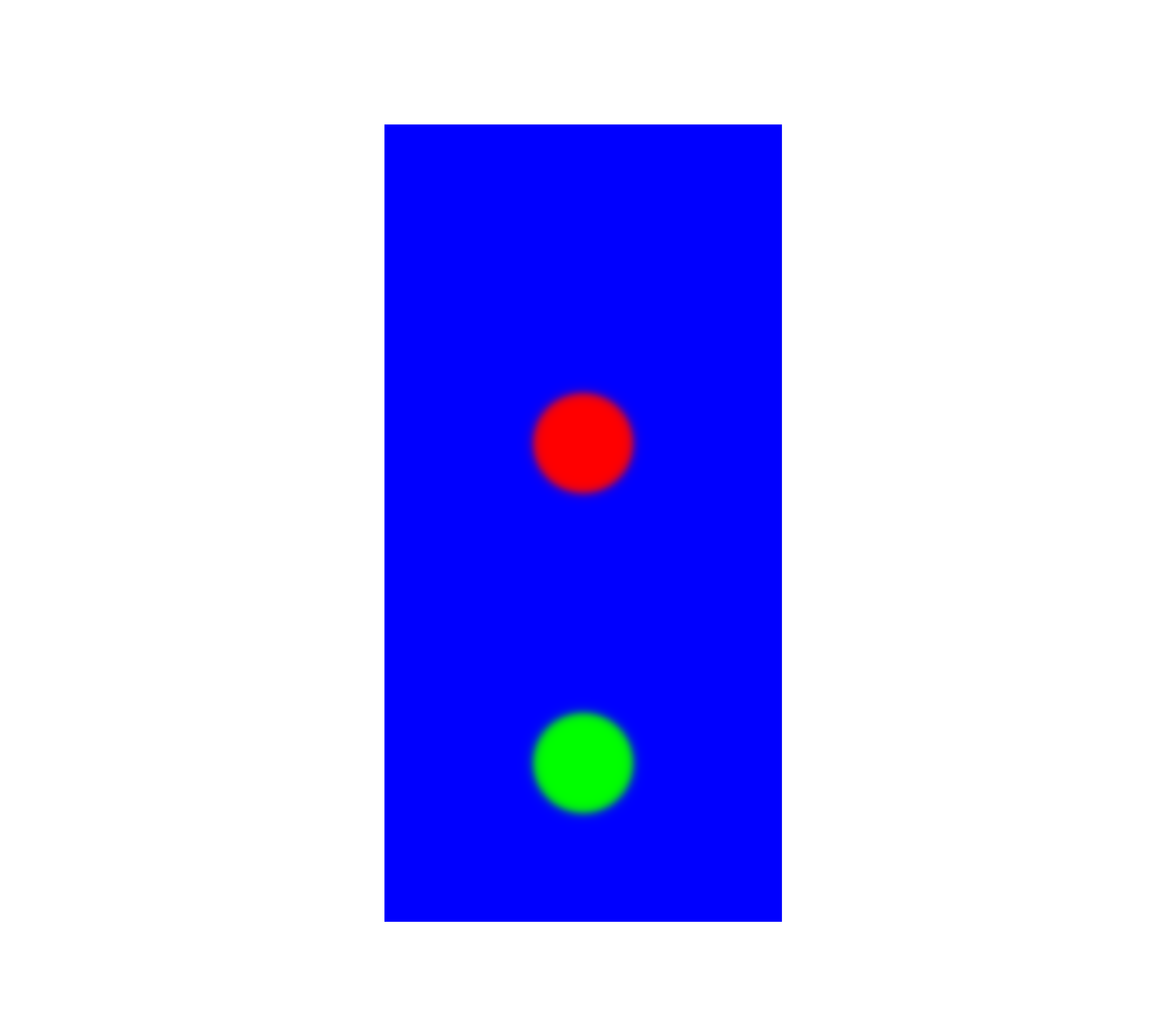}
    \caption{Initial condition of three-component fluid flows.}
    \label{fig1}
\end{figure}

In this section, we examine the effect of mobility in our simulations. We set our variable and constant mobility as in Eq.~\eqref{eq:mobility}. As can be seen, the $C_i(1-C_i)$ term, which was applied in \cite{kim2007numerical}, is multiplied by the constant mobility introduced in \cite{hagh}. In fact, instead of a constant mobility applied to the whole domain, a variable mobility is implemented on the interfaces only. Figure \ref{fig1} shows initial condition of three components so that red, green, and blue colors correspond to component 1, 2, and 3, respectively. Green bubble rises because of buoyant force to stick to the red droplet. This situation, bubble-droplet rising, is a critical situation in our simulation which can cause mass loss. Table \ref{table1} shows properties of the three-component fluid flow system in a $200\times400$ computational domain we investigate.

\begin{table}[h]
	\centering
	\begin{tabular}{ l c c c c c c c }
		\hline
		\hline
		$\rho_1$ & $\rho_2$ & $\rho_3$  & g & $\frac{\tau_1}{\tau_3}=\frac{\tau_2}{\tau_3}$& $\sigma_{12}=\sigma_{13}=\sigma_{23}$   \\
		$1$ & $0.01$ & $1$ &  $10^{-6}$ & $2$ & $10^{-4}$ \\
		\hline
		\hline
	\end{tabular}
	\caption{Physical properties of three-component fluid flow system}
	\label{table1}
\end{table}

\begin{figure}[H]
    \centering
   \includegraphics[width= 1\textwidth,trim={0cm 0cm 0cm 0cm},clip]{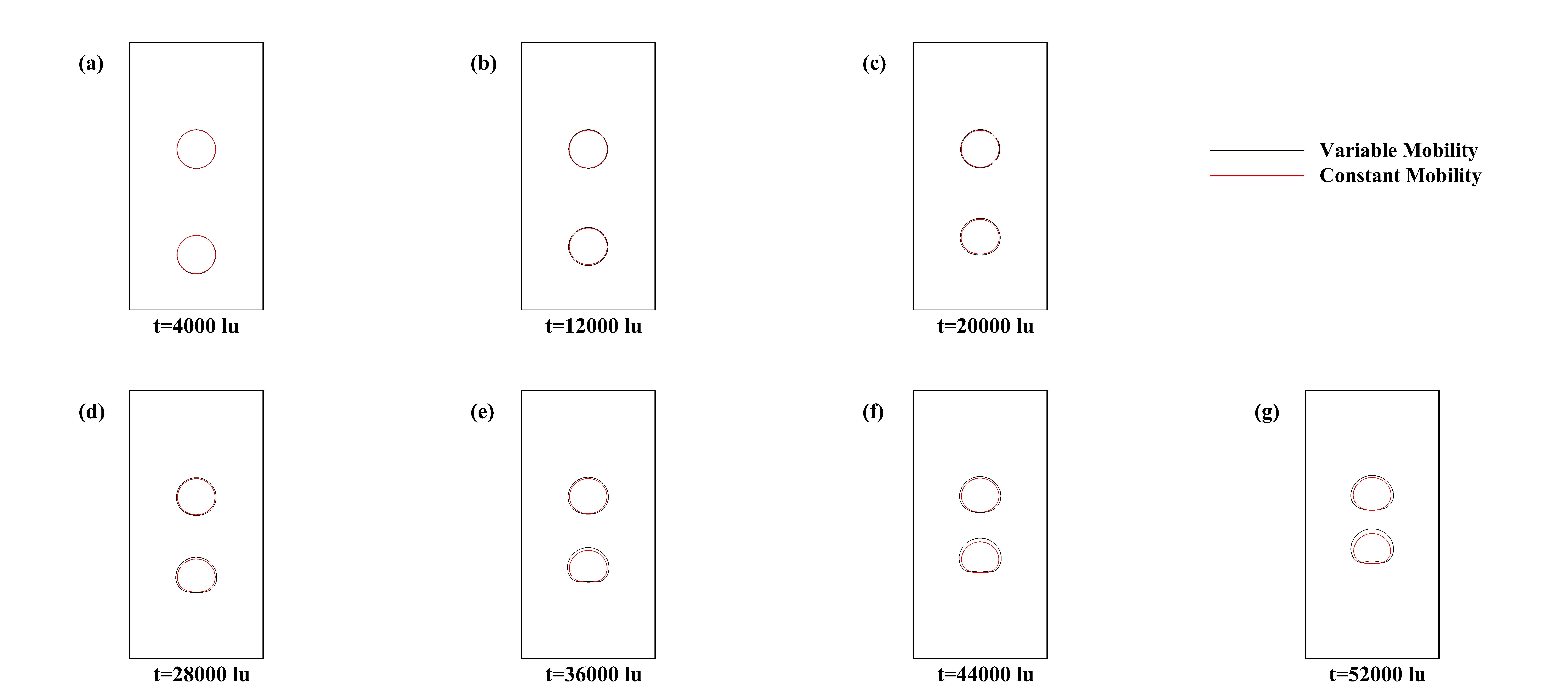}
    \caption{Line contour of bubble and droplet with constant and variable mobility}
    \label{fig2}
\end{figure}

Seven line contours at different times were plotted to compare our simulation using variable mobility and constant mobility. As shown in Fig. \ref{fig2}, red lines and black lines represent level contour of 1 of components obtained by constant mobility and variable mobility, respectively. The droplet, which is under no buoyant force, preserves its mass with variable mobility while with constant mobility its mass is dissipated. Therefore, it can be seen in the Fig. \ref{fig2} that red line droplet shrink inside of the black line droplet over time. Also, the rising bubble loses its mass using constant mobility, so the black line bubble get more buoyant force, indicating that it deforms sooner and rises faster than red line bubble.   

\begin{figure}[H]
    \centering
   \includegraphics[width= 1\textwidth,trim={0cm 0cm 0cm 0cm},clip]{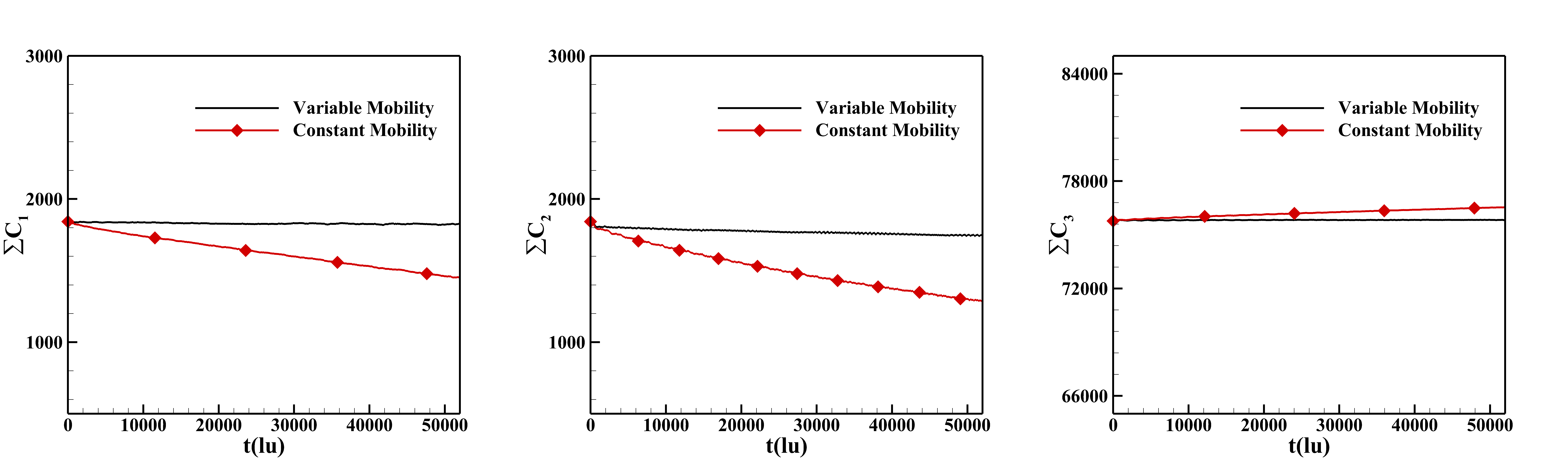}
    \caption{Variation of total concentration of each component over time.}
    \label{fig3}
\end{figure}

To compare our results precisely, concentration of each component is recorded when it is above 0.5. As Fig. \ref{fig3} shows, summation of each $C_i$ in all of the lattices demonstrates that green bubble and red droplet gradually lose their mass, and the losing mass adds to the blue component mass. When the number of lattices are increased, transformations of mass between components decrease with constant mobility while variable mobility can omit this transformation even in a small domain. Since variable mobility conserves mass exceedingly, we perform our simulations using variable mobility in the present study.

\subsection{Single phase power-law fluid flows between two parallel plates}
\begin{figure}[H]
    \centering
   \includegraphics[width= 0.85\textwidth,trim={0cm 0cm 0cm 0cm},clip]{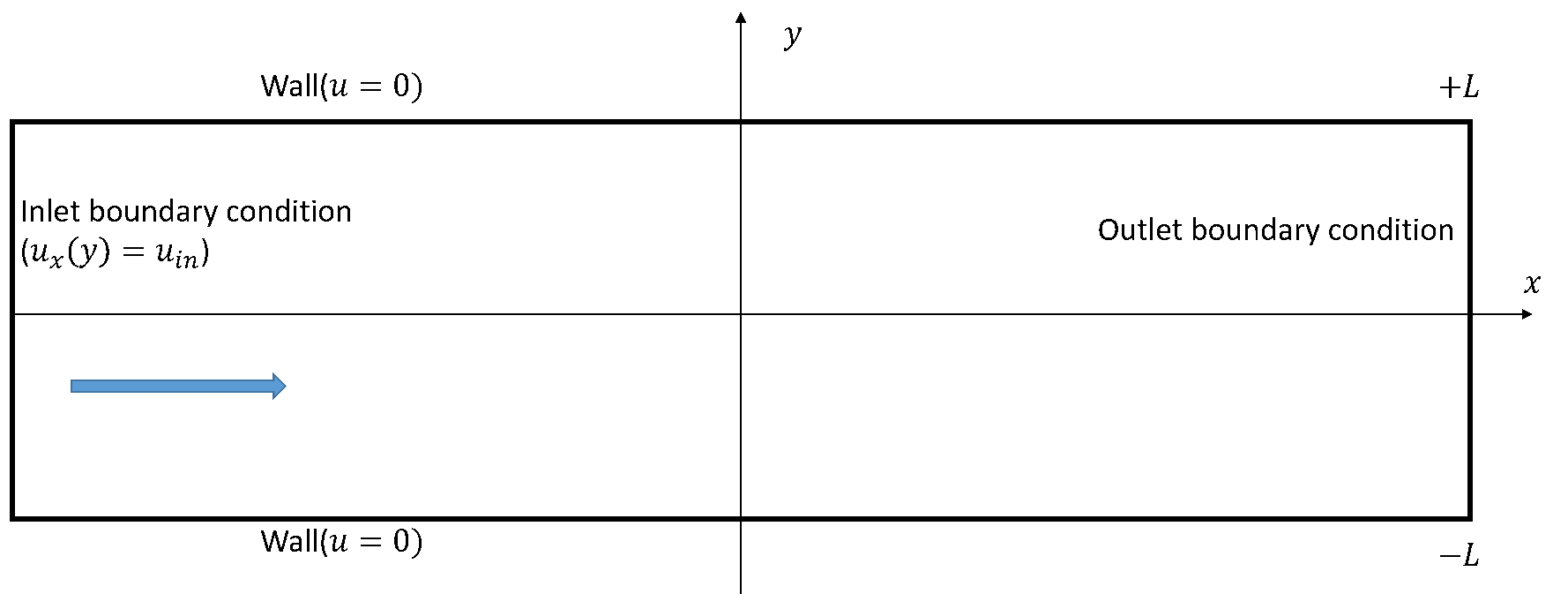}
    \caption{Schematic of a single phase power-law fluid flow between two parallel plates.}
    \label{fig4}
\end{figure}

In this subsection and the following one, two benchmarks are conducted to validate the implementation of our code. First benchmark is a pressure-driven flow between two fixed parallel plates at a distance of $2L$ in the $y$-direction while fluid flow moves in the $x$-direction as shown in Fig. \ref{fig4}.

\begin{figure}[H]
    \centering
   \includegraphics[width= 1\textwidth,trim={0cm 0cm 0cm 0cm},clip]{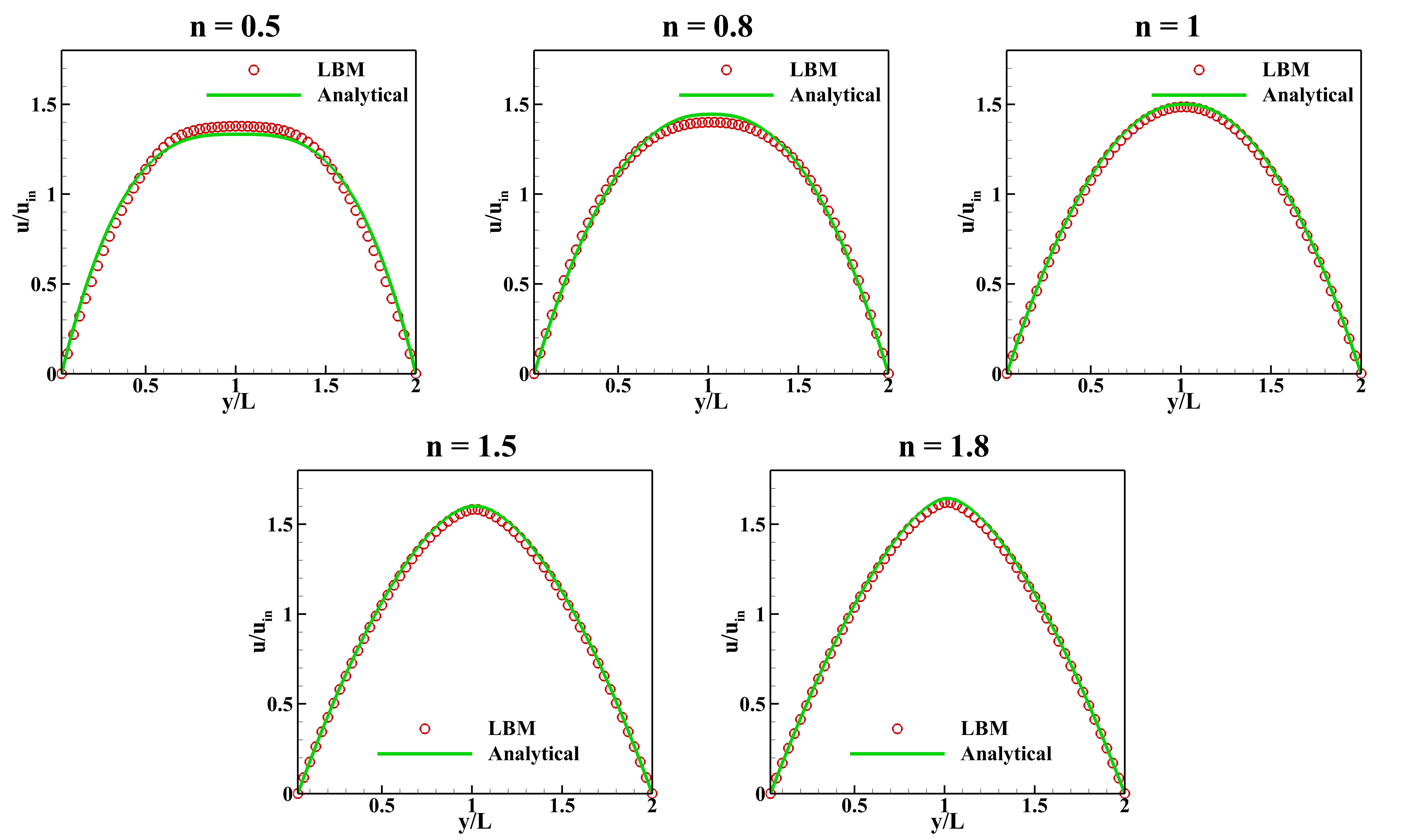}
    \caption{Comparison of analytical and LBM velocity profiles in a 2D channel for power-law fluids with indices of $n=0.5, 0.8, 1, 1.5, 1.8$.  }
    \label{fig5}
\end{figure}

The analytical solution for steady state velocity profile with power-law index is as follows: \cite{BATTISTELLA2020104249}
\begin{equation}
    u_x(y)=u_{in}\frac{2n+1}{n+1}(1-{\lvert \frac{y}{L}\rvert}^{\frac{n+1}{n}})
\end{equation}
where $u_{in}$ is the inlet velocity, $y$ is the distance from the center of two parallel plates, and $n$ is the power-law index. The simulations are performed for various power-law indices including shear-thickening, shear-thinning, and Newtonian fluids. The No-slip boundary condition is imposed on the top and bottom walls. Velocity inlet \cite{velocity} and convective boundary conditions \cite{outflow} are applied on the left and right, respectively. The simulations are performed in a two dimensional rectangular domain where width of the channel is 60 and the length is 240 lattice unit. Initially, compositions of $C_1$ and $C_2$ are equal to zero so that the channel is filled completely by the third component. A constant velocity in the $x$-direction ($u_{in}=0.1$) is imposed at the left boundary. Several simulations by various power-law indices are carried out and theoretical and numerical velocity profiles are compared at steady state as shown in Fig. \ref{fig5}. 

The relative $L_2$-norm of the velocity has been calculated at steady state condition between analytical and numerical results as presented in the following equation:
\begin{equation} \label{eq:ltwonorm}
    E_{rel}=\frac{\lvert \lvert u_x - u^{analytical}_x \rvert \rvert}{\lvert \lvert u^{analytical}_x \rvert \rvert}
\end{equation}
where $u_x$ is the $x$ component of velocity, $u^{analytical}_x$ is the $x$ component of analytical velocity, and $E_{rel}$ is the relative $L_2$-norm of the velocity. 

Table \ref{table2} shows the relative errors of the power-law fluid velocity between two parallel plates at different $n$. It can be concluded that the simulation results are in good agreement with the analytical solution.

\begin{table}[h]
	\centering
	\begin{tabular}{ l c c c c c c c }
		\hline
		\hline
		n & 0.5 & 0.8  & 1 & 1.5 & 1.8   \\
		$E_{rel}$ & $0.03$ & $0.02$ &  $0.01$ & $0.01$ & $0.01$ \\
		\hline
		\hline
	\end{tabular}
	\caption{Relative errors of the velocity at different power-law indices.}
	\label{table2}
\end{table}

\subsection{Two-phase pressure driven power-law fluid flows between two parallel plates}
\begin{figure}[H]
    \centering
   \includegraphics[width= 1\textwidth,trim={0cm 0cm 0cm 0cm},clip]{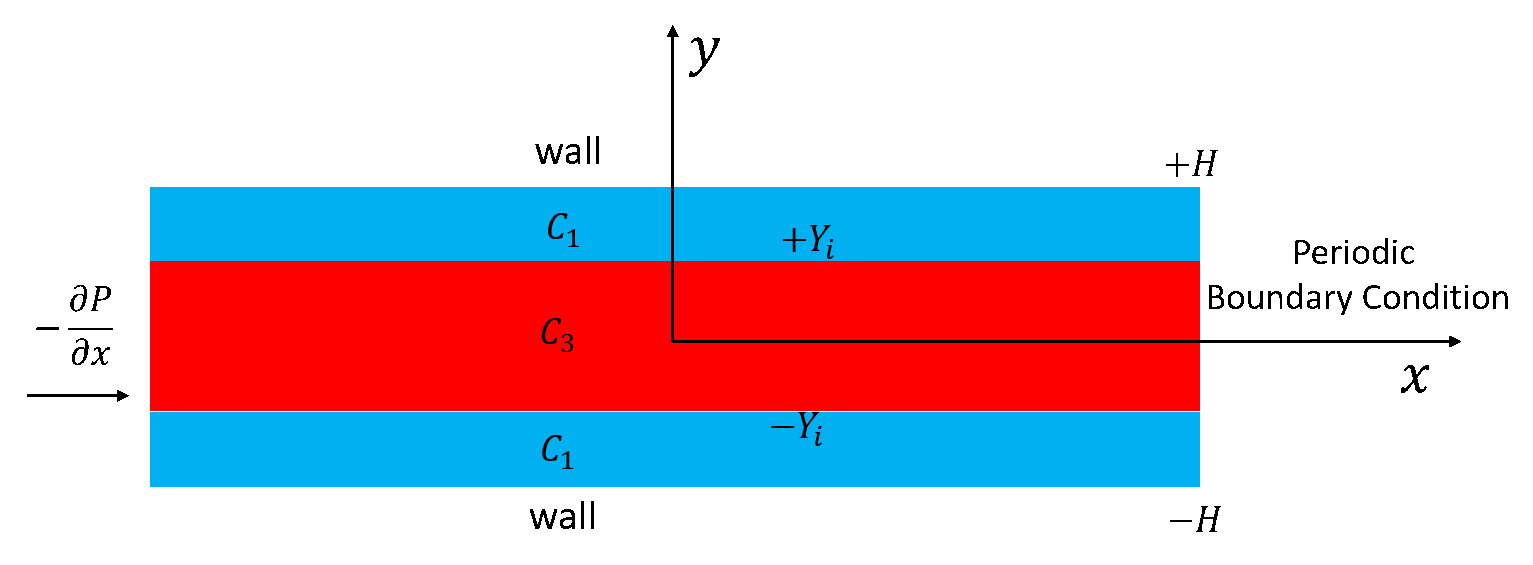}
    \caption{Schematic of the two-phase layered fluid flow between two parallel plates. Component 3 is placed at the center, component 2 is equal to zero, and component 1 is placed at top and bottom.   }
    \label{fig:fig6}
\end{figure}

In this subsection, two-phase layered flow is considered between two parallel plates which is driven by pressure gradient. This benchmark is used to assess the ability of the present LB model in phases interaction. So, one of the component is set to zero initially to obtain a two-phase flow. As illustrated schematically in Fig.~\ref{fig:fig6}, each component fills half of the 2D channel and the composition 2 is set to zero. The height of the channel is $2H$ and the component 3 is placed at $\lvert y\rvert < Y_i$. Periodic boundary condition is applied on the left and right boundaries while bounce-back scheme is imposed on the top and bottom. Each fluid is driven by a constant body force  ($-\frac{\partial P  }{\partial x}=F_b$) whose value is $1.5\times10^{-8}$. The simulation is performed in a domain with $5\times200$ lattice units. Density ratio is equal to unity but different viscosity ratios are investigated. The viscosity ratio is defined as $\lambda={\mu_3}/{\mu_1}$ which is used for Newtonian systems. For non-Newtonian systems new parameter is defined as $\lambda^p = \mu^p_3 / \mu^p_1$ which is used instead of $\lambda$ \cite{haihuliu}. Truncated power-law model is applied in the simulation and the relaxation parameter limit, $\tau_{min}$ and $\tau_{max}$, are set. The surface tension is $\sigma_{13}=5\times 10^{-4}$  and the interface thickness is $\xi=4$. The simulation is run until it reaches the steady state which is determined by $\frac{{\lvert \lvert u(x,t) - u(x,t-\delta t) \rvert \rvert}_2}{{\lvert \lvert u(x,t) \rvert \rvert}_2}<10^{-8}$. As discussed in \cite{haihuliu}, the analytical velocity in the $x$-direction is obtained as follows: 
\begin{equation}
 u^*_x(y) = \left\{ \begin{array}{rcl}
\frac{n_R}{n_R +1}(-\frac{1}{\mu^R_P}\frac{\partial P}{\partial x})^{\frac{1}{n_R}}(Y^{\frac{1+n_R}{n_R}} _i - {\lvert y \rvert}^{\frac{1+n_R}{n_R}}) + 
\frac{n_B}{n_B +1}(-\frac{1}{\mu^B_P}\frac{\partial P}{\partial x})^{\frac{1}{n_B}}(Y^{\frac{1+n_B}{n_B}} _i - {\lvert y \rvert}^{\frac{1+n_B}{n_B}}) & \mbox{if} &  \lvert y \rvert < Y_i \\
\frac{n_B}{n_B +1}(-\frac{1}{\mu^B_P}\frac{\partial P}{\partial x})^{\frac{1}{n_B}}(Y^{\frac{1+n_B}{n_B}} _i - {\lvert y \rvert}^{\frac{1+n_B}{n_B}}) & \mbox{otherwise} & 
\end{array}\right.
\end{equation}
It should be noted that the analytical maximum velocity is $u^*_x(y=0)$. The maximum velocity should not be greater than $10^{-2}$ because of compressibility error \cite{haihuliu}.

\begin{figure}[H]
    \centering
   \includegraphics[width= 1\textwidth,trim={0cm 0cm 0cm 0cm},clip]{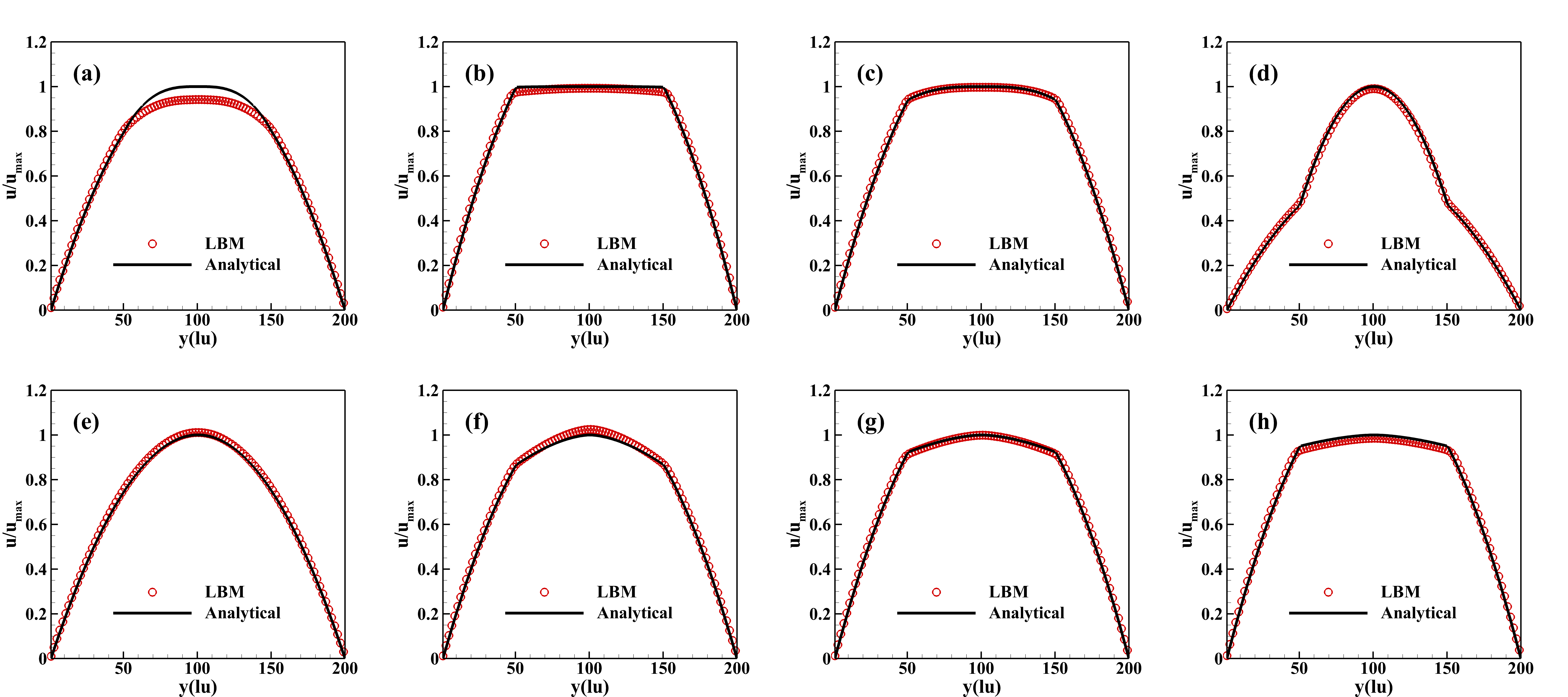}
    \caption{Comparison of analytical and LBM velocity profiles in a 2D channel for two-phase power-law layered flow at different $\lambda^P$ and power-law indices. (a) $n_3=0.5$, $n_1=1$, $\lambda^P=0.002$, (b) $n_3=0.5$, $n_1=1$, $\lambda^P=0.02$, (c) $n_3=0.5$, $n_1=1$, $\lambda^P=0.004$, (d) $n_3=1$, $n_1=1$, $\lambda=0.3$, (e) $n_3=1$, $n_1=1$, $\lambda=1$, (f) $n_3=1.5$ ,$n_1=1$, $\lambda^P=2000$, (g) $n_3=1.5$, $n_1=1$, $\lambda^P=5000$, and (h) $n_3=1.5$, $n_1=1$, $\lambda^P=10000$.    }
    \label{fig7}
\end{figure}

Several simulations with different $\lambda^P$ and power-law indices are performed. Analytical and numerical velocity profiles are compared as shown in Fig.  \ref{fig7}.
The relative $L_2$-norm which was presented in Eq.~\eqref{eq:ltwonorm} is obtained in each simulation to investigate the accuracy of this model. Table \ref{table3} shows the relative errors of the simulated and analytical velocity. As it can be seen in the table \ref{table3}, $\lambda^P$  and $\lambda$ are used for non-Newtonian and Newtonian systems, respectively.

\begin{table}[h]
    \centering
    \begin{tabular}{ l c c c c c c  }
    \hline
    \hline
         $\lambda^P$ & $\lambda$ & $n_3$  & $n_1$ & $E_{rel}$ \\ 
    \hline
   $0.002$ &  & $0.5$ &  $1$ & $0.04$ \\
   $0.02$ &  & $0.5$ &  $1$ & $0.01$ \\
   $0.004$ &  & $0.5$ &  $1$ & $0.01$ \\
    & $0.3$ & $0.5$ &  $1$ & $0.01$ \\
    & $1$ & $1$ &  $1$ & $0.01$ \\
   $2000$ &  & $1$ &  $1$ & $0.01$ \\
   $5000$ &  & $1.5$ &  $1$ & $0.01$ \\
   $10000$ &  & $1.5$ &  $1$ & $0.01$ \\
    \hline
    \hline
    \end{tabular}
    \caption{Relative errors of the velocity for different power-law indices and different $\lambda^P$ and $\lambda$  in the two-phase power-law layered flow driven by pressure gradient.}
    \label{table3}
\end{table}
The relative errors are in order of $10^{-2}$ in each simulation which shows that the numerical results are in good  agreement with the analytical solution. It should be noted that as long as the maximum velocity decreases, the compressibility error will be also smaller, therefore, the relative errors are lower. 

\section{Results}
\subsection{Drag coefficient and dimensionless numbers of a bubble rising in a liquid}
In this section, the terminal velocity of a bubble rising in a power-law fluid is investigated. When the bubble rising (subscript $g$) in a liquid (subscript $l$) reaches its terminal velocity, buoyancy and drag forces exerting on the bubble become equal. Therefore, the drag coefficient can be written as \cite{20cad03008e541e78c2c7e51aa97e6e7, lima}:
\begin{equation}
C_D = \frac{4dg(\rho_l - \rho_g)}{3\rho_l U^2 _t}
\end{equation}

The Reynolds number, Morton number, and E\"{o}tv\"{o}s number in power-law fluids are defined as \cite{lima, CHHABRA199089}

\begin{equation}
    Re=\frac{\rho_l U^{2-n}d^n _e}{k}
\end{equation}

\begin{equation}
    Mo = \frac{g\Delta \rho}{\rho^2 _l \sigma^3}[k{(\frac{U}{d_e})}^{n-1}]^4
\end{equation}
\begin{equation}
   Eo  = \frac{(\rho_l - \rho_g)gd^2 _e}{\sigma}
\end{equation}
Where $d_e$ is the equivalent diameter which is defined as $d_e=\sqrt[3]{\frac{6V_b}{\pi}}$ using bubble total volume, $V_b$. In the Newtonian fluids, the drag coefficient is only a function of Reynolds number, but in the power-law fluids the index $n$ has also effect on the drag coefficient ($C_D=f(Re,n)$)\cite{CHHABRA199089}.

Li \etal \cite{lima} proposed a drag correlation for a bubble rising in a shear-thinning liquid, considering deformation of bubble and rheological properties of fluids besides Reynolds number. They showed that the proposed correlation is capable of handling spherical and deformed bubble.
\begin{equation}
    C_D=\frac{16}{Re}(1+0.43Re^{0.44})(1+3.868n^{0.7528}(1-E)^{0.6810}).
\end{equation}
where $E$ is the aspect ratio of the bubble, expressing the ratio of vertical and horizontal diameters of the bubble. They showed that this correlation have great agreement with experimental results under the finite range of $0.05<Re<300$. It should be noted that in our study, a bubble rising by buoyant force, the bubble experiences $E\geq1$ because the only force exerted on the bubble is from bubble rear. But if a bubble rises in other non-Newtonian liquids applying forces from both left and right like viscoelastic fluids, bubble aspect ratio can be less than unity. 

\subsection{Grid independency}

\begin{figure}[H]
	\centering
	\includegraphics[width= 0.85\textwidth,trim={0cm 0cm 0cm 0cm},clip]{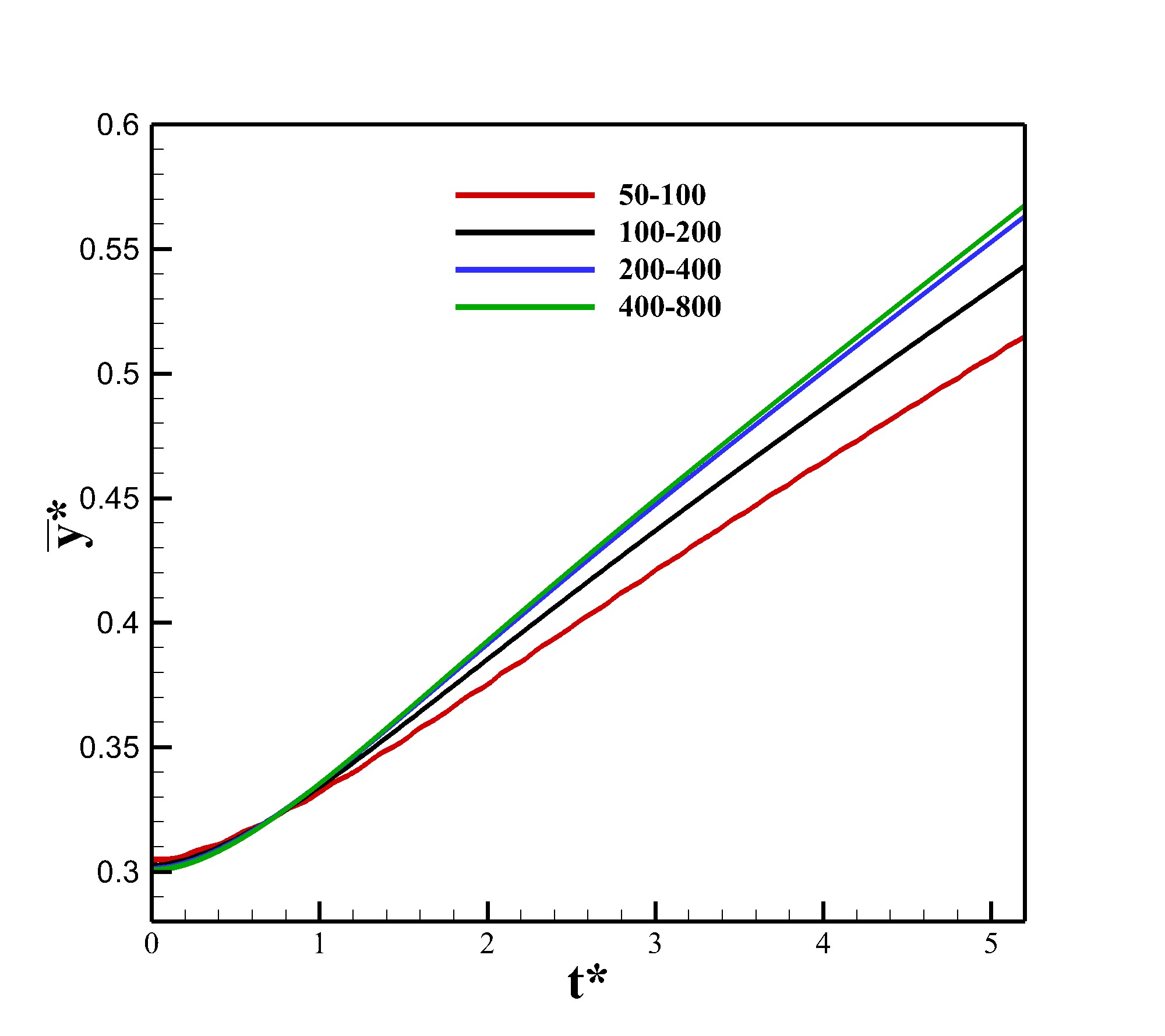}
	\caption{Gravity center of the bubble as function of time for different grid resolution.   }
	\label{fig8}
\end{figure}

Grid independency test in four domains is performed to find the grid independent computational domain. A domain of $ L\times2L$ is considered which $L$ is the length of it. The no-slip boundary condition is imposed on the top and bottom and periodic boundary condition is applied on the left and right boundaries. Time evolution of gravity center of the single bubble, which is under buoyant force, is plotted for each grid in Fig. \ref{fig8}. Table \ref{table4} shows rheological properties of liquid and physical properties of gas-liquid system, where $\tau_g$, $d_g$, and $g$ represent relaxation time of the disperesed component, the rising bubble, diameter of the bubble, and gravitational acceleration, respectively.

\begin{table}[h]
	\centering
	\begin{tabular}{ l c c c c c c c c c}
		\hline
		\hline
		Grid number& Domain Size & $\rho_l / \rho_g $  & $k$ & $n$& $\tau_g$& g & $d_g$& $\sigma$ \\ 
		\hline
		$1$ & $50\times100$ &  $800$ & $2\times10^{-4}$ & $0.5$ & $0.5$ & $8\times10^{-6}$& $\frac{2L}{5}$& $1.044\times10^{-4}$\\
		$2$ &  $100\times200$ &  $800$ & $10^{-4}$ & $0.5$ & $0.5$ & $10^{-6}$& $\frac{2L}{5}$& $5.22\times10^{-5}$ \\
		$3$ & $200\times400$ &  $800$ & $5\times10^{-5}$ & $0.5$ & $0.5$ &$1.25\times10^{-7}$ & $\frac{2L}{5}$& $2.61\times10^{-5}$ \\
		$4$ & $400\times800$ & $800$ &  $2.5\times10^{-5}$ & $0.5$ & $0.5$& $1.56\times10^{-8}$ &$\frac{2L}{5}$& $1.3\times10^{-5}$\\
		\hline
		\hline
	\end{tabular}
	\caption{Rheological and physical properties of different computational domain.}
	\label{table4}
\end{table}

The parameters used in the Table \ref{table4} are selected so that dimensionless numbers, Reynolds, E\"{o}tv\"{o}s, and Morton numbers, remain constant in each grid. The gas-liquid system is assumed to be quiescent initially and initial condition for concentration is as follows:
\begin{equation}
    \phi_d(\bm{x},0)=0.5-0.5 \tanh(\frac{\sqrt{(x-0.5 L)^2 + (y-0.6 L)^2}-0.5{d_b}}{0.5{\xi}})
\end{equation}

In this study, the velocity of the bubble and gravity center of the bubble are recorded in average-form as follows:
\begin{equation}
    \Bar{V}=\frac{\sum C_g V}{\sum C_g}
\end{equation}
\begin{equation}
    \Bar{y}=\frac{\sum C_g y}{\sum C_g}
\end{equation}
Also, we define dimensionless form for velocity and location for better comparison by dividing them to $\sqrt{gd_g} $ and $2L$, respectively.
 To choose a suitable domain, dimensionless velocity and center of gravity of the bubble are examined. Magnitude of dimensionless terminal velocity of the bubbles are $0.198$, $0.243$, $0.268$, and $0.273$ for grid 1, grid 2, grid 3, and grid 4, respectively. Since the error between grid 3 and grid 4 is less than $1\%$, grid 3 is suitable for the rest of our simulations. Also, curves correspond to the grid 3 and 4 are so close as shown in Fig. \ref{fig8}

\subsection{ Bubble rising in a shear-thinning liquid}

In this section, the terminal velocity of a bubble in a shear-thinning liquid, whose viscosity decreases by applying shear stress, is investigated.
To investigate prediction of correlation proposed by Li \etal \cite{lima}, some numerical simulation with different rheological properties are conducted using present LB model. All the simulations represent shear-thinning behavior of the liquid ($n<1$). It should be noted that the liquid contains solely a bubble and is completely pure.
Properties of gas-liquid systems were provided in table \ref{table5}

\begin{table}[H]
    \centering
    \begin{tabular}{l c c c c c c c}
    \hline
    \hline
        Case& $\rho_c / \rho_d $  & $n$ & $k$& $g$ &$\tau_d$& $d_b$& $\sigma$ \\ 
    \hline
   $1$ & $800$ &  $0.5$ & $1\times10^{-4}$ & $5\times10^{-7}$ & $0.5$ & $\frac{L}{4}$& $2.61\times10^{-6}$\\ \hline
   
   $2$ & $800$ &  $0.5$ & $1\times10^{-4}$ & $5\times10^{-7}$ & $0.5$ & $\frac{L}{4}$& $2.61\times10^{-5}$\\ \hline
   
   $3$ & $800$ & $0.5$ & $2\times10^{-4}$ & $5\times10^{-7}$ & $0.5$ & $\frac{L}{4}$ & $5\times10^{-4}$  \\ \hline
   
   $4$ & $800$ & $0.6$ & $5\times10^{-4}$ & $2\times10^{-7}$ & $0.5$ & $\frac{L}{4}$ & $2.61\times10^{-5}$  \\ \hline
   
   $5$ & $800$ & $0.63$ & $5\times10^{-4}$ & $2\times10^{-7}$ & $0.5$ & $\frac{L}{4}$ & $2.61\times10^{-5}$  \\ \hline
  
   $6$ & $800$ &  $0.8$ & $1\times10^{-2}$ & $5\times10^{-7}$ & $0.5$ & $\frac{L}{4}$& $2.61\times10^{-5}$\\
   
    \hline
    \hline
    \end{tabular}
    \caption{Physical properties of the gas-liquid systems.}
    \label{table5}
\end{table}

Since the behavior of the bubble is determined using dimensionless numbers, the Reynolds and E\"{o}tv\"{o}s numbers for each case are shown in table \ref{table6}:
\begin{table}[H]
    \centering
    \begin{tabular}{ l c c c}
    \hline
    \hline
    Case & $Re $   & $Eo$ \\ \hline
    $1$ & $4.05 $   & $478.9$ \\  \hline
    $2$ & $4.29 $   & $47.89$ \\  \hline
    $3$ & $1.36 $   & $2.5$ \\  \hline
    $4$ & $0.44 $  & $19.15$ \\  \hline
    $5$ & $0.99 $  & $19.15$ \\  \hline
    $6$ & $0.47 $   & $47.89$ \\

    \hline
    \hline
    \end{tabular}
    \caption{Dimensionless numbers at different cases.}
    \label{table6}
\end{table}

\begin{figure}[H]
    \centering
   \includegraphics[width= 1\textwidth,trim={0cm 0cm 0cm 0cm},clip]{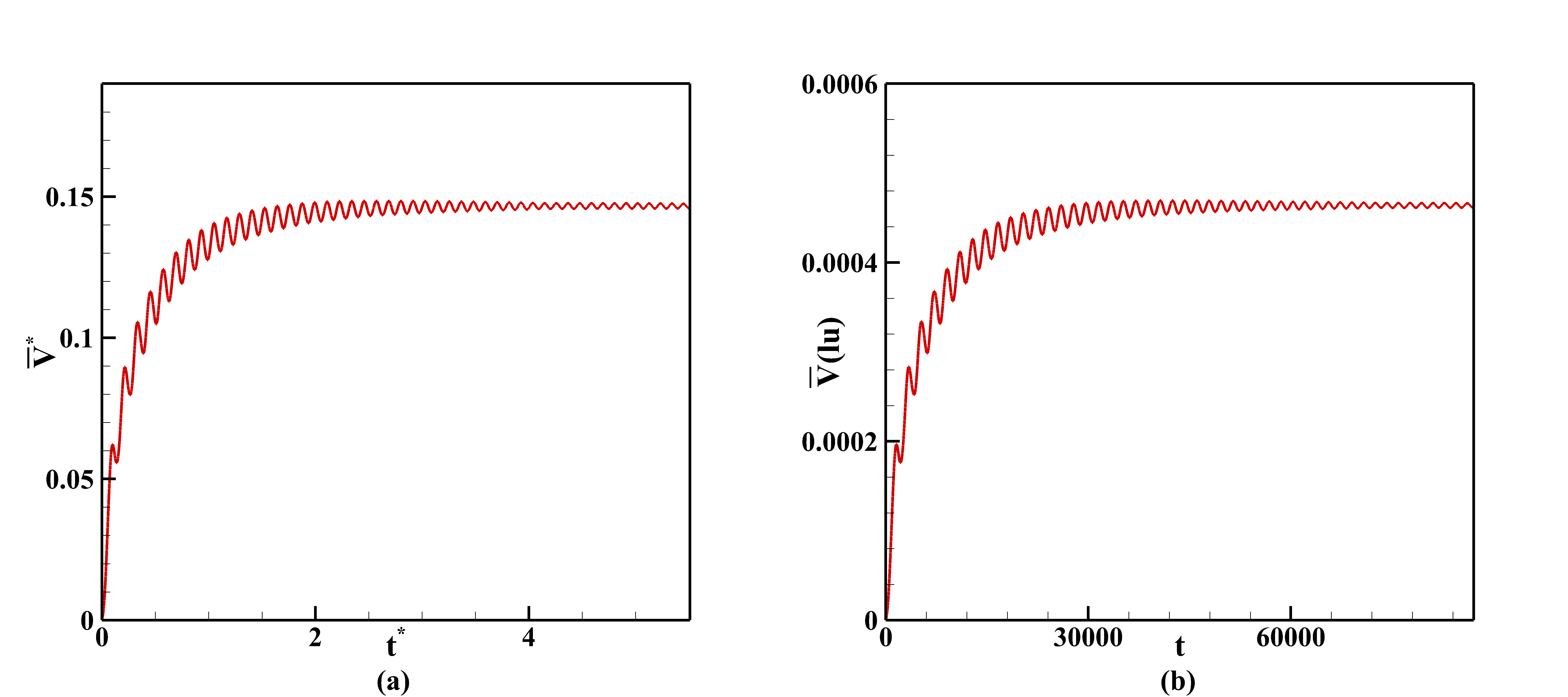}
    \caption{Dimensionless (a) and lattice unit (b) velocity of the bubble in the case 4}
    \label{fig9}
\end{figure}

Velocity of the bubble increases gradually and reaches a constant value (steady state). To show this process, for instance, in case 4, lattice unit and dimensionless velocity of the bubble are shown in Fig. \ref{fig9} (a) and (b), respectively.

\begin{figure}[H]
    \centering
   \includegraphics[width= 0.85\textwidth,trim={0cm 0cm 0cm 0cm},clip]{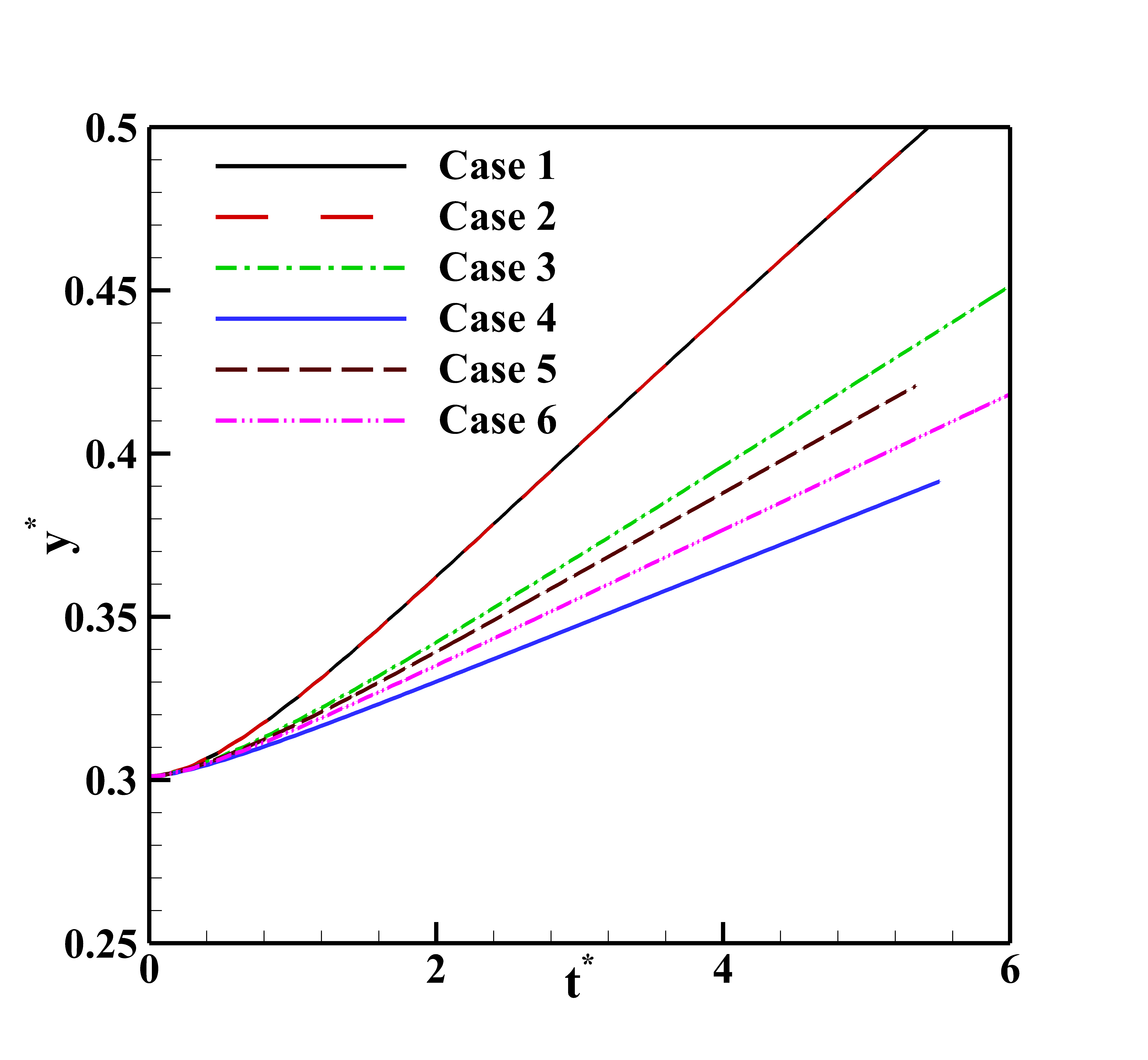}
    \caption{Dimensionless gravity center of the bubble in six cases.}
    \label{fig10}
\end{figure}

As drag coefficient is dependent on the Reynolds number, the Reynolds number plays an important role in determining the location of the bubble. The effect of Reynolds number is shown in Fig. \ref{fig10} by recording the gravity center of the bubble, which was made dimensionless by the height of the domain, for all the cases. Initially, the bubble starts to move gradually, demonstrating that the slope of lines in all the cases is equal to zero. Then, the slope of line increases to reach a constant value. At higher Reynolds number, the slope of curves and gravity center are higher, indicating that the bubble associated with the higher Reynolds number goes upward faster.

To compare our results with the mentioned correlation, a relative error, $(|numerical-correlation|/correlation)$, for each simulation is provided in Table \ref{table7} with the maximum relative error of 19$\%$, minimum relative error of 0, and average relative error of 5.6$\%$. Table \ref{table7} shows that terminal velocity of the bubble at different conditions are in good agreement with empirical correlation proposed by Li  \etal \cite{lima}. One of the reasons behind small deviation between the results is that in the present model, the shear-thinning liquid is assumed to be pure, unlike experimental results which were done using carboxymethyl
cellulose (CMC) solutions.

\begin{table}[H]
    \centering
    \begin{tabular}{ l c c c c c c c c c}
    \hline
    \hline
     Case & $n$ & $E$& $U_t$ (Li et. al \cite{lima}) &$U_t$ (present)& Relative Error \\ 
    \hline
   $1$ & $0.5$ &  $0.77$ & $0.00159$ & $0.00156$ & $0.01$\\ \hline
   $2$ & $0.5$ &  $0.73$ & $0.00159$ & $0.00157$ & $0.01$\\ \hline
   $3$ & $0.5$ &  $0.98$ & $0.0012$ & $0.0011$ & $0.08$\\ \hline
   $4$ & $0.6$ &  $0.96$ & $0.00046$ & $0.00046$ & $0$\\ \hline
   $5$ & $0.63$ &  $0.9$ & $0.0006$ & $0.00063$ & $0.05$\\ \hline
   $6$ & $0.8$ &  $0.94$ & $0.00071$ & $0.00085$ & $0.19$\\

    \hline
    \hline
    \end{tabular}
    \caption{Comparison of terminal velocity obtained by present LB model and correlation.}
    \label{table7}
\end{table}

Different shapes of bubble are obtained at different Reynolds and E\"{o}tv\"{o}s numbers. Since surface tension tries to maintain minimum surface area of the bubble, higher E\"{o}tv\"{o}s number represents more deformation tendency of the bubble. Also, higher Reynolds number represents higher momentum force in the liquid, resulting in higher exerted force on the bubble. To analyze better, concentration, viscosity, and velocity contours are plotted at steady state condition for each case.

\begin{figure}[H]
    \centering
   \includegraphics[width= 1\textwidth,trim={0cm 0cm 0cm 0cm},clip]{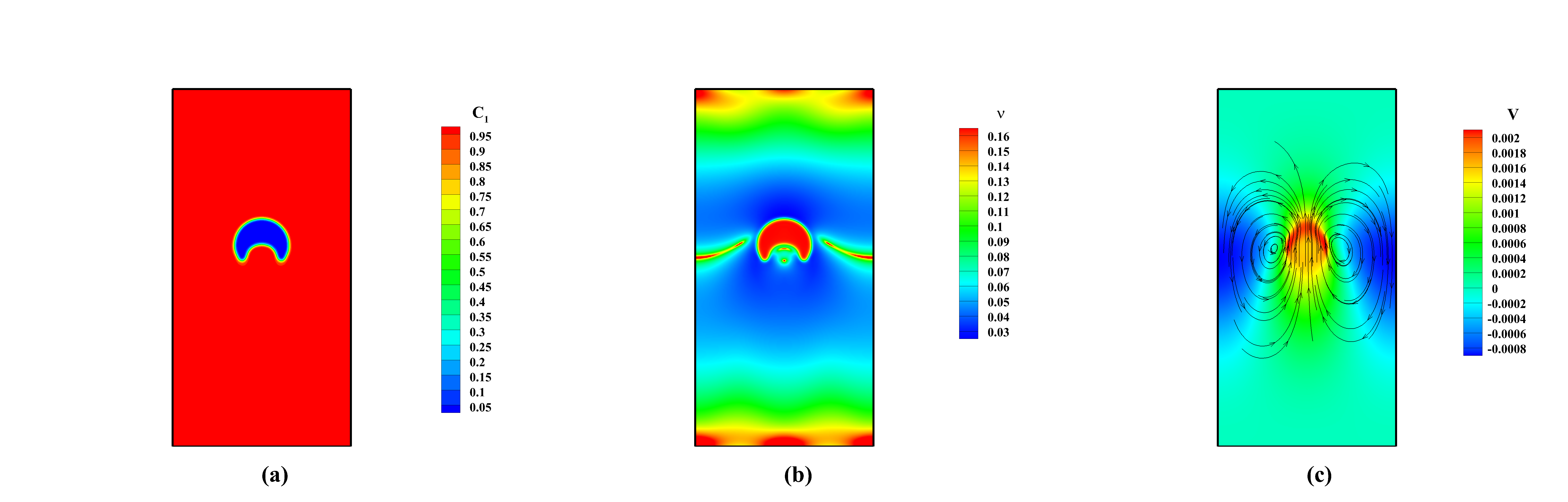}
    \caption{Final (a) concentration, (b) kinematic viscosity, and (b) velocity contours of case 1. }
    \label{fig11}
\end{figure}

In the case 1, the surface tension is very low; therefore, the bubble tends to deform more as shown in Fig. \ref{fig11} (a). Kinematic viscosity distribution of the liquid shows that when the bubble rises, viscosity decreases around the bubble because of the shear-thinning behavior. As bubble rises, shear rate around the bubble increases, leading to decrease in viscosity. The farther away from the bubble, the higher the viscosity. When the bubble gets far from the bottom boundary, the viscosity in the vicinity of the boundary increases or when the bubble gets close to the top boundary, the viscosity nearby the top boundary becomes lower gradually which indicate shear-thinning behavior of the liquid. Also, left and right boundaries are periodic, which indicate less shear rate in the left and right of the bubble, causing a thin line of high viscosity in the left and right of the bubble.

\begin{figure}[H]
    \centering
   \includegraphics[width= 1\textwidth,trim={0cm 0cm 0cm 0cm},clip]{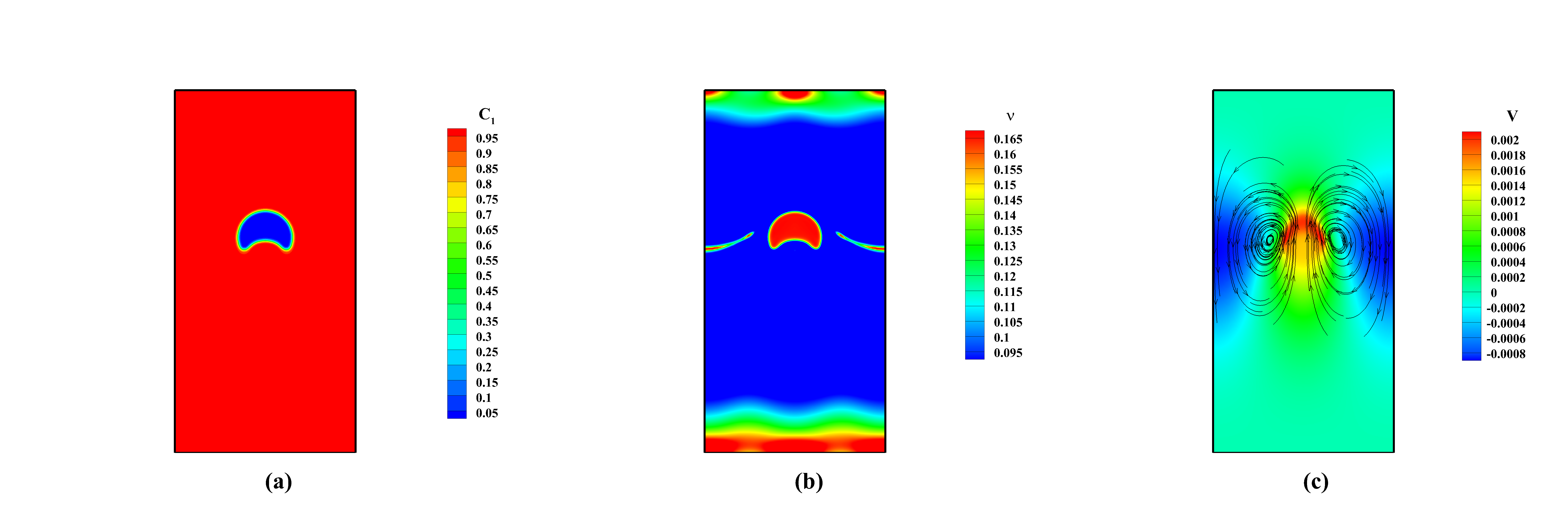}
    \caption{Final (a) concentration, (b) kinematic viscosity, and (b) velocity contours of case 2. }
    \label{fig12}
\end{figure}

In the case 2, the physical properties are the same as the case 1 but the surface tension. Therefore, the final shape of the bubble in the case 2, which has more surface tension, is less deformed with respect to the final shape of the bubble in the case 1 as shown in Fig. \ref{fig12} (a). Also, Fig. \ref{fig12} (b) demonstrates that by increasing surface tension, the kinematic viscosity around the bubble increases because a low deformation indicates a low shear rate.

\begin{figure}[H]
    \centering
   \includegraphics[width= 1\textwidth,trim={0cm 0cm 0cm 0cm},clip]{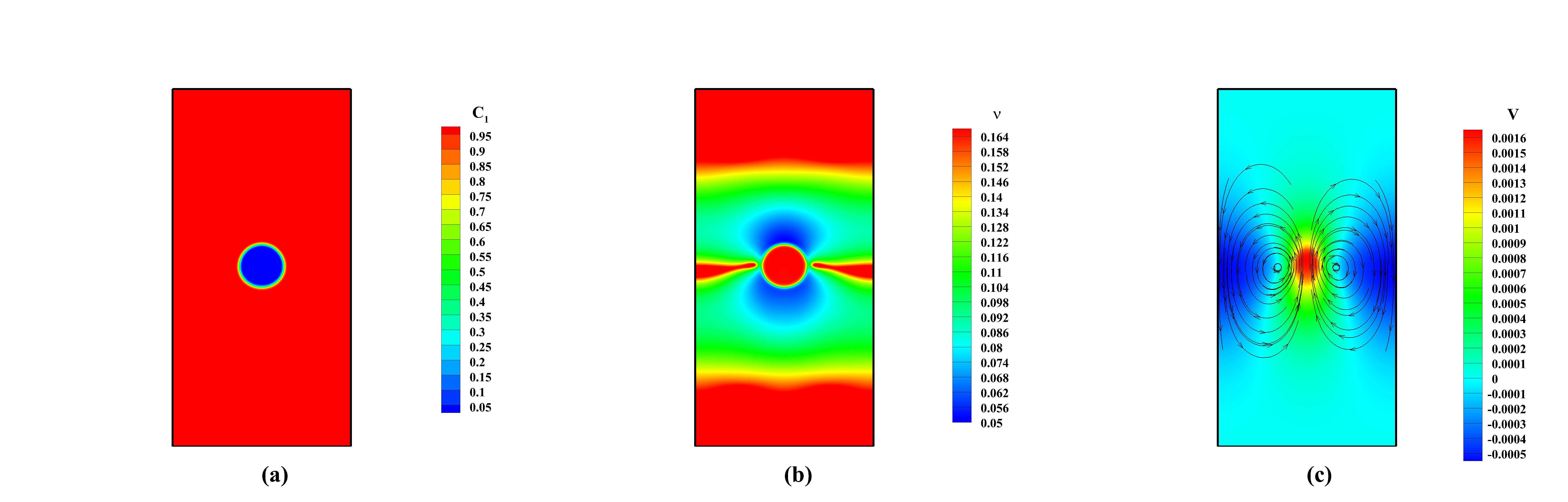}
    \caption{Final (a) concentration, (b) kinematic viscosity, and (b) velocity contours of case 3. }
    \label{fig13}
\end{figure}

In the case 3, Reynolds and E\"{o}tv\"{o}s numbers decrease by increasing consistency index and surface tension coefficient, respectively. As a result, the bubble is capable of maintaining spherical shape over time as shown in Fig. \ref{fig13} (a). Also, Fig. \ref{fig13} (b) shows that kinematic viscosity distribution is higher with respect to the cases 1 and 2. Also, the two regions on the left and right of the bubble, having large viscosity, is bigger with respect to the cases 1 and 2.

\begin{figure}[H]
    \centering
   \includegraphics[width= 1\textwidth,trim={0cm 0cm 0cm 0cm},clip]{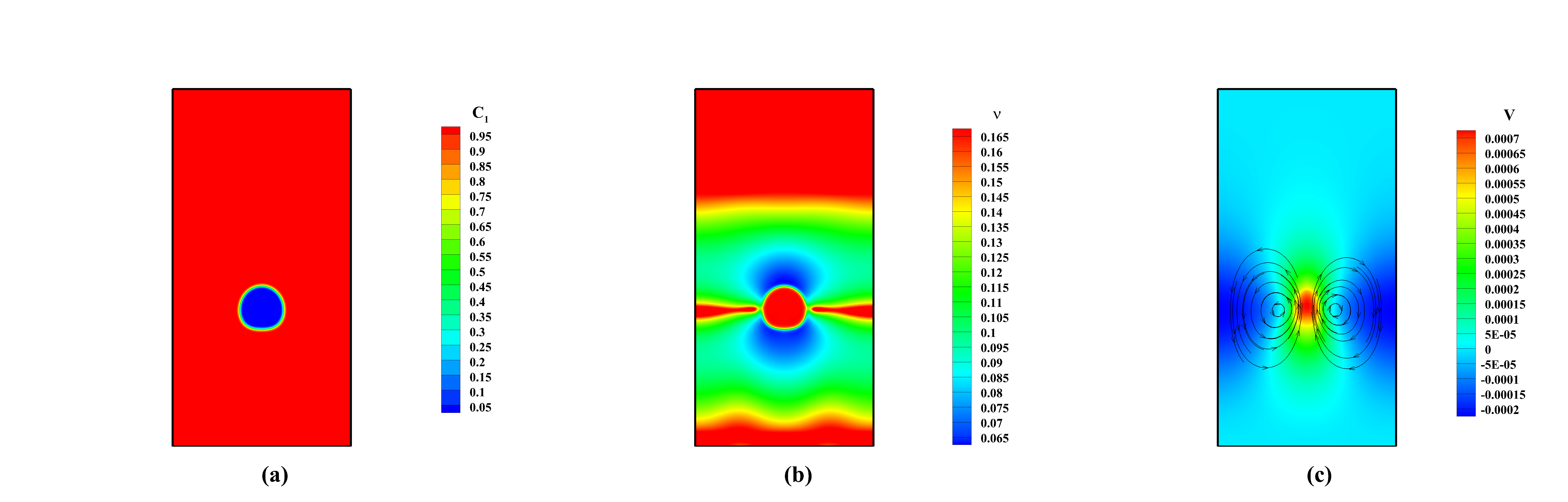}
    \caption{Final (a) concentration, (b) kinematic viscosity, and (b) velocity contours of case 4. }
    \label{fig14}
\end{figure}

\begin{figure}[H]
    \centering
   \includegraphics[width= 1\textwidth,trim={0cm 0cm 0cm 0cm},clip]{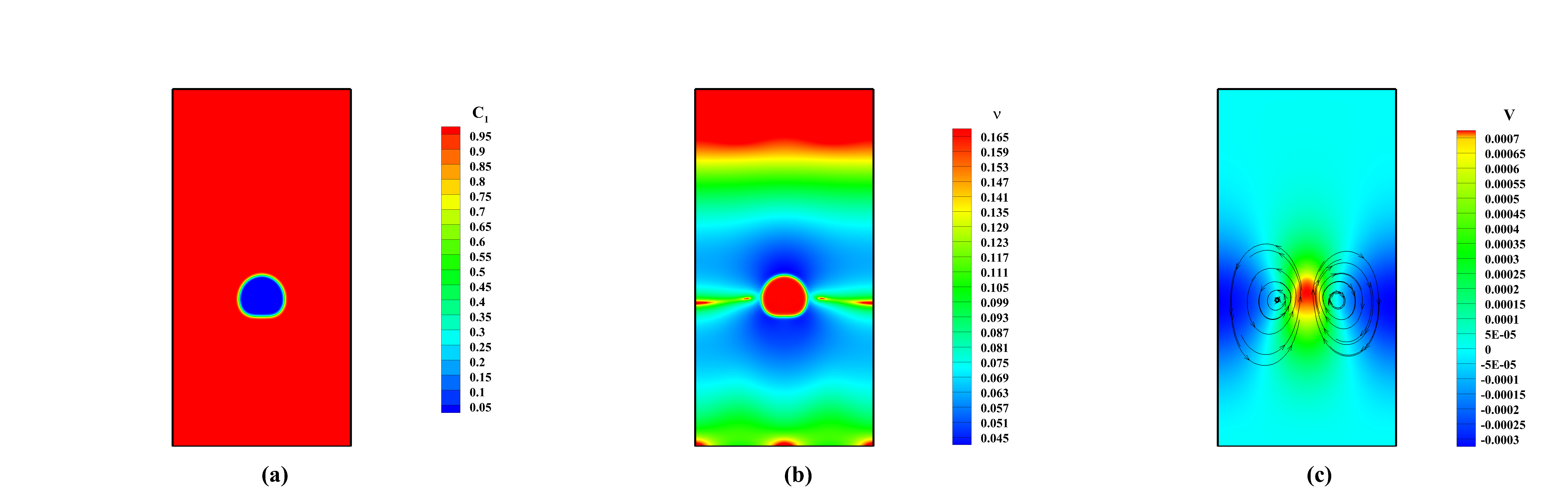}
    \caption{Final (a) concentration, (b) kinematic viscosity, and (b) velocity contours of case 5. }
    \label{fig15}
\end{figure}

In the cases 4 and 5, higher power-law indices, $n=0.6$ and $n=0.63$, are implemented while all other parameters are the same. Since in our simulation the shear rate magnitude is less than unity, shear thinning behavior increases by increasing power-law index \cite{ioannou2017droplet}. Therefore, as Fig. \ref{fig14} (b) and Fig. \ref{fig15} (b) shows, $n=0.63$ exhibits more shear-thinning characteristics than $n=0.6$. In addition, due to the kinematic viscosity distribution in the domain, the Reynolds number in the case 5 is approximately two times larger than that in the case 4.

\begin{figure}[H]
    \centering
   \includegraphics[width= 1\textwidth,trim={0cm 0cm 0cm 0cm},clip]{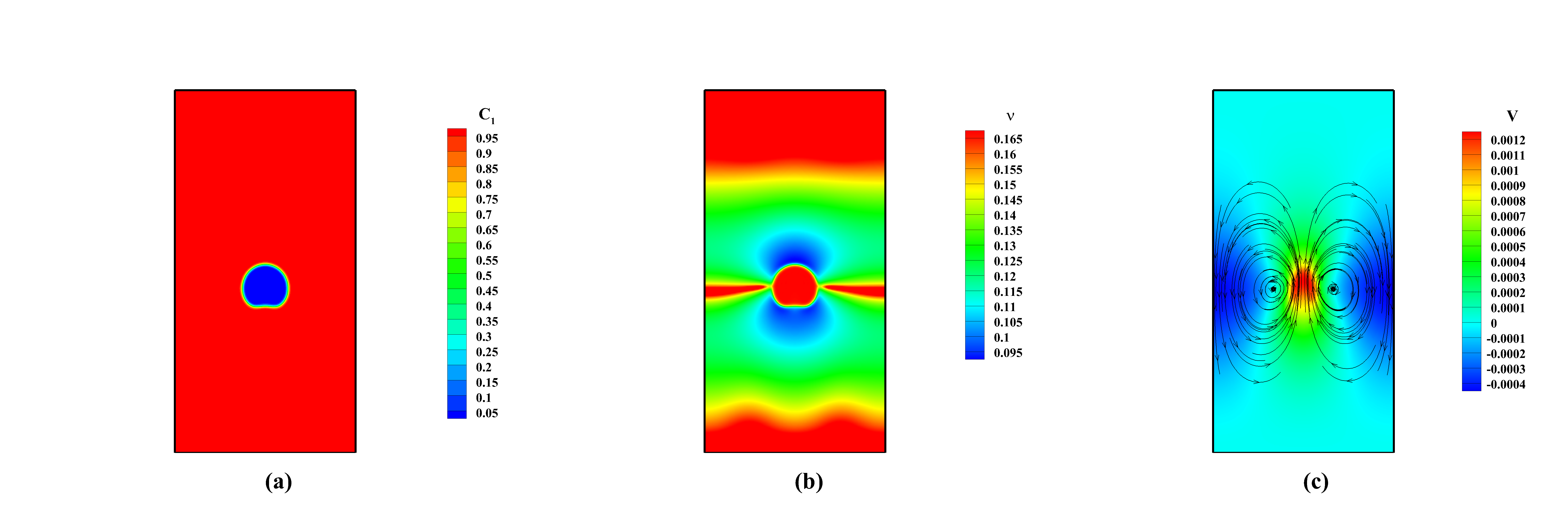}
    \caption{Final (a) concentration, (b) kinematic viscosity, and (b) velocity contours of case 6. }
    \label{fig16}
\end{figure}

In some cases, for instance case 6, as shown in Fig. \ref{fig16} (b), two regions at the top and bottom have constant, high kinematic viscosity because these regions experience minimum and equal shear rate at specific times. When the bubble rises, it exerts large shear stress around itself, resulting in a decrease in the viscosity of surrounding shear-thinning fluid. Therefore, the bottom region retrieve its high, constant kinematic viscosity when the bubble take away from it, and this constant-viscosity region get larger as long as the bubble goes upward. As a result, when the bubble gets close to the top boundary, shear rate becomes variable and larger at the top region, transforming the top region to a small, high-viscosity region.

The rising bubble pushes the surrounding fluid upward, forcing it to circulate. As a result, two wakes appear, experiencing low velocity as shown in Fig. \ref{fig11} (c)- \ref{fig16} (c), which make the bubble rear flattened or skirted, especially in high E\"{o}tv\"{o}s numbers.

\section{Conclusions}

In this paper, a power-law scheme was implemented in a LB model to capture both shear-thinning and Newtonian behavior of a fluid. A variable mobility was introduced in this study to reduce interface dissipation between phases. A comparison between constant and variable mobility in a three-component system was presented, indicating that the variable mobility can preserve mass of components better.

Two cases were investigated to test the accuracy of the model. A single phase power-law fluid flow driven by a constant inlet velocity between two parallel plates was simulated to show the small relative error between the numerical and analytical fully-developed velocity profile. Also, a two-phase power-law fluid flow driven by pressure between two parallel plates was investigated at different power-law indices and viscosity ratios to show the small relative errors between numerical simulation and analytical solution of velocity profiles. The results showed the strength of the present model in simulating power-law and Newtonian fluids with wide range of power-law indices.

The dynamics of a rising bubble in the shear-thinning liquid was examined numerically using a LB model. Results show that the rising bubble deforms at high E\"{o}tv\"{o}s number and reaches a steady state shape and velocity. Also, the bubble moves upward faster and has higher location at high Reynolds number. The terminal velocity obtained by an experimental drag coefficient was compared to the simulated terminal velocity, demonstrating that the results of the present study are in good agreement with the experiment. The average, maximum, and minimum relative errors for six cases were $5.6\%$, $19\%$, and $0$, respectively. When the bubble rises in a shear-thinning liquid, the viscosity around the bubble decreases because of the increase of shear rate. Therefore, the viscosity nearby the bubble, which experiences the highest shear rate, is the lowest and it increases gradually when gets farther away from the bubble. Also, the regions which are too far from the bubble have constant and maximum viscosity. The velocity contour and streamlines show that the rising bubble generates two wakes around itself, containing low or negative velocity. 

In future study, we will use multiple relaxation time (MRT) to capture higher viscosity ratios. Also, higher Reynolds numbers should be captured to investigate wider range of dimensionless numbers. 

\section*{Conflict of interest}
The authors declare no conflict of interest.

\section*{Data Availability}
The data used to generate the results in the manuscript are available from the corresponding author upon reasonable request.

\newpage
\bibliographystyle{elsarticle-num}
\bibliography{bibliography}

\end{document}